\documentclass[twocolumn,epjc3]{svjour3}  
\smartqed  
\RequirePackage{graphicx}
\usepackage[utf8]{inputenc}
\usepackage{graphics}
\usepackage{epstopdf}
\usepackage{xcolor}
\usepackage{rotating}
\usepackage{dblfloatfix} 
\usepackage{microtype} 

\usepackage{amsmath,amssymb}
\usepackage[hypertexnames,setpagesize,%
    pdftex,%
    colorlinks,%
    citecolor=blue,%
    hyperindex,%
    plainpages=false,%
    bookmarksopen,%
    bookmarksnumbered%
  ]{hyperref} 
  
\usepackage[numbers,square,comma,sort&compress]{natbib}

\usepackage{textcomp}

\usepackage{lineno,xcolor}
\usepackage{tikz}
\usetikzlibrary{arrows,shapes,positioning}

\usepackage{multirow}

\journalname{Eur. Phys. J. C}

\begin{document}


\title{
Alpha backgrounds in the AMoRE-Pilot experiment 
}

\author{V.~Alenkov\thanksref{addr1}
\and
H.W.~Bae\thanksref{addr2}
\and
J.~Beyer\thanksref{addr3}
\and
R.S.~Boiko\thanksref{addr4}
\and
K.~Boonin\thanksref{addr5}
\and
O.~Buzanov\thanksref{addr1}
\and
N.~Chanthima\thanksref{addr5}
\and
M.K.~Cheoun\thanksref{addr6}
\and
S.H.~Choi\thanksref{addr7}
\and
F.A.~Danevich\thanksref{addr4}
% ...
}
\thankstext{corrauthor1}{e-mail: ejjeon@ibs.re.kr}
\thankstext{corrauthor2}{e-mail: na.liansha@gmail.com}
\institute{JSC FOMOS-Materials, Moscow 107023, Russia \label{addr1}
\and
Department of Physics, Kyungpook National University, Daegu 41566, Korea \label{addr2}
\and
Physikalisch-Technische Bundesanstalt, D-38116 Braunschweig, Germany \label{addr3}
\and
Institute for Nuclear Research of NASU, 03028 Kyiv, Ukraine \label{addr4}
\and
Nakhon Pathom Rajabhat University, Nakhon Pathom 73000, Thailand \label{addr5}
\and
Department of Physics, Soongsil University, Seoul 06978, Korea \label{addr6}
\and
Department of Physics and Astronomy, Seoul National University, Seoul 08826, Korea \label{addr7}
% ...
}

%
%



\date{Received: date / Accepted : date}

\maketitle

\begin {abstract}
The Advanced Mo-based Rare process Experiment (AMoRE)-Pilot experiment is an initial phase of the AMoRE search for neutrinoless double beta decay of $^{100}$Mo, with the purpose of investigating the level and sources of backgrounds.  
Searches for neutrinoless double beta decay generally require ultimately low backgrounds.
Surface $\alpha$ decays on the crystals themselves or nearby materials can deposit a continuum of energies that can be as high as the $Q$-value of the decay itself and may fall in the region of interest (ROI). 
To understand these background events, we studied backgrounds from radioactive contaminations internal to and on the surface of the crystals or nearby materials with Geant4-based Monte Carlo simulations.
In this study, we report on the measured $\alpha$ energy spectra fitted with the corresponding simulated spectra for six crystal detectors, where sources of background contributions could be identified through high energy $\alpha$ peaks and continuum parts in the energy spectrum for both internal and surface contaminations.
We determine the low-energy contributions from internal and surface $\alpha$ contaminations by extrapolating from the $\alpha$ background fitting model.
\end{abstract}

	

\section{Introduction}
\label{sec:intro}
One of the most fundamental questions about neutrinos is whether they are classified as Majorana or Dirac particles. The only practical method to answer this question is to find neutrinoless double beta decays (0$\nu\beta\beta$), which would establish neutrinos as Majorana particles.
The rate of 0$\nu\beta\beta$ (which is the inverse of the mean lifetime of this process) is proportional to the effective Majorana neutrino mass squared, which is a function of the masses and mixing angles of the three neutrinos and the so-called unknown Majorana phases~\cite{DellOro:2016tmg,Mohapatra2007,Patrignani2016,Giunti2007}. 
The most sensitive lower limits on half-lives of 0$\nu\beta\beta$ decay for different isotopes such as $^{76}$Ge, $^{82}$Se,  $^{100}$Mo, $^{130}$Te, and $^{136}$Xe are at the level of $10^{24}$ to $10^{26}$ years~\cite{gando2016,agostini2020,abgrall2014,alvis2019,anton2019,adams2020,azzolini2019,arnold2021}.

AMoRE-Pilot is the initial phase of the AMoRE project that aims at searching for the neutrinoless double beta decay of $^{100}$Mo~\cite{Bhang2012,Alenkov2015}. 
It consists of an array of six low-temperature phonon-scintillation detectors based on calcium molybdate crystals~($^{48depl}$Ca$^{100}$MoO$_{4}$, CMO) with a total mass of 1.9~kg~\cite{alenkov2019}.
The set-up operated at the Yangyang Underground Laboratory (Y2L) from 2015 to 2018. The data taken between Aug 10 and Dec 20, 2017, with a total exposure of 0.457~kg~yr, are used in this analysis.

One of the AMoRE-Pilot goals is to establish a better understanding of the background conditions to improve the level of 0$\nu\beta\beta$ half-life sensitivity for future versions of AMoRE. 

The ultimate sensitivity of AMoRE is limited by the internal radioactive contamination of the CMO detectors by traces of uranium and thorium with their daughters~\cite{luqman2017}.
In addition, contaminations on the crystals surfaces or nearby materials can deposit energies in the active volume of  the detector~\cite{bucci2009,johnson2012,patavina,alessandria,yu2021,vacri2021}, which could possibly appear in the experimental region of interest (ROI)
that is $\pm10$~keV around the $^{100}$Mo $Q$-value of 3.034~MeV, estimated based on the energy resolution of the detector~\cite{gbkim2017}.
Although $\alpha$ events from these backgrounds can be separated from double beta decay signals by particle identification (PID) methods in AMoRE~\cite{gbkim2016,ikim2017}, misidentification of  $\beta/\gamma$ events can occur induced by the surface contamination from natural decay chains can leave residual backgrounds; hence, it is essential to understand these residual backgrounds in depth.

Contributions from radioactive $\alpha$-decays can be identified by their high-energy peaks and continuum parts in the background spectrum in internal contaminations and surface contaminations of crystals and nearby materials, respectively. For this,  measured $\alpha$ energy spectra are modeled by fitting them with simulated $\alpha$ energy spectra for each crystal. 
The background modeling for $\beta/\gamma$ spectra will be described in a separate work. 

The paper is structured as follows: the AMoRE-Pilot experimental setup is described in Section 2. Section 3 describes the internal background measurements of CMO crystals. Section 4 describes background simulations for both internal and surface contaminations. Section 5 provides details regarding the $\alpha$ spectra modeling developed by fitting the measured $\alpha$ spectra with the simulated data and gives details about the results, including background contributions in the ROI as well as low-energy contributions. Finally, conclusions are provided in Section 6. 

\section{AMoRE-Pilot experimental setup}
\label{sec:2}
\begin{figure*}[ht]
\centering
\resizebox{0.9\textwidth}{!}{
\includegraphics{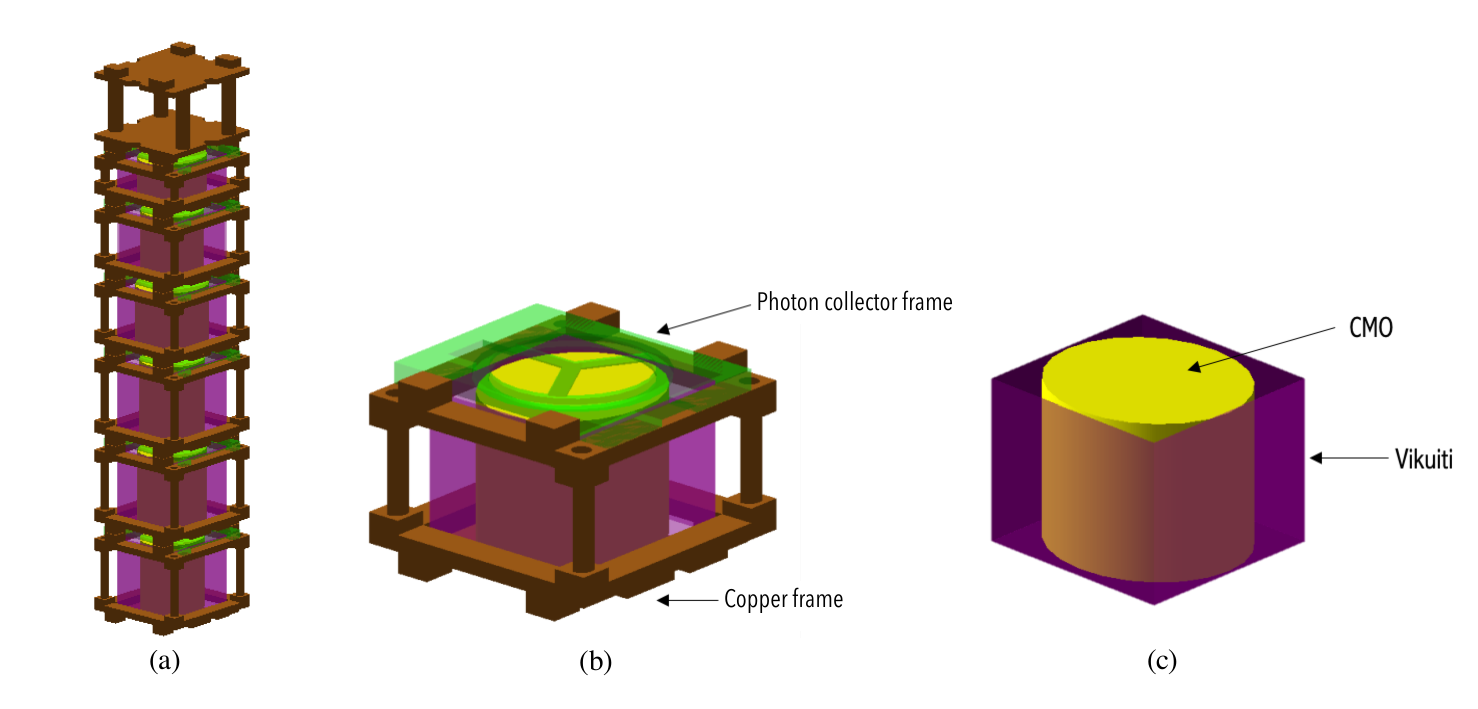}
}
\caption{AMoRE-pilot detector setup: 
(a) detector array; (b) photon frame location; (c) Vikuiti reflector location.} \label{fig:pict1}
\label{fig:detector}
\end{figure*}
\begin{table*}[ht]
\begin{center}
\caption{CMO crystal scintillator sizes, masses, and surface areas:}
\label{tab:detector}
\begin{tabular}{c  c  c  c  c  c }
\hline
Crystal & \multicolumn{2}{c}{Diameter (cm)} & Height (cm) & Mass (g) & Surface area (cm$^2$)\\
      &Minor axis & Major axis & & \\ \hline
      Crystal 1 & 4.30     & 5.25   &   2.6    & 196      & 74.56 \\
      Crystal 2   & 4.09   & 4.60 	& 4.1		& 256 	& 84.88 \\
      Crystal 3 & 4.79   & 5.38	& 4.2		& 352	& 106.83 \\
      Crystal 4 & 4.27   & 4.68	& 5.2		& 354 	& 104.53 \\
      Crystal 5 & 4.37   & 5.14	& 5.1		& 390	& 111.59 \\
      Crystal 6 & 4.04 & 4.84 & 5.1	& 340	& 101.97 \\ \hline
    \end{tabular} 
\end{center}
\end{table*}
\begin{table}[ht]
\begin{center}
\caption{Vikuiti reflector masses and surface areas}
\label{tab:vikuiti}
\begin{tabular}{ccc}
\hline
Reflector number & Mass (g) & Surface area (cm$^2$)\\
 & &  \\ \hline
      reflector$\_$1 & 0.47   & 106.23   \\
      reflector$\_$2 & 0.64   & 143.50 	\\
      reflector$\_$3 & 0.67   & 146.07	\\
      reflector$\_$4 & 0.67   & 173.05	\\
      reflector$\_$5 & 0.67   & 170.48	\\
      reflector$\_$6 & 0.67   & 170.48	\\ \hline
    \end{tabular}
    \end{center} 
\end{table}
The AMoRE-pilot detector assembly consists of an array of six crystals stacked vertically with a total mass of 1.9~kg, as shown in Fig.~\ref{fig:detector}(a). The six crystal detectors are labeled as Crystal 1 (C1) to Crystal 6 (C6).
The CMO crystals are produced from calcium, which is depleted in $^{48}$Ca (less than 0.001\% to reduce the background from the 2$\nu\beta\beta$ decay of $^{48}$Ca~\cite{annenkov2008}) and molybdenum enriched in $^{100}$Mo (more than 95\%~\cite{kang2017}) by JSC “Fomos Materials”~\cite{fomos} in the framework of R\&D with the AMoRE collaboration. All the CMO scintillation crystals have been pulled by the Chochralski technique using  the so-called double crystallization procedure to reduce the level of the impurities,  except for Crystal 2, which was grown by a single crystal pulling procedure.
Each crystal has an elliptical cylinder shape with a mass ranging from 196 g to 390 g. The details for each crystal are listed in Table~\ref{tab:detector}. Crystals are assembled inside a highly radiopure  copper frame with high electrical conductivity, as shown in Fig.~\ref{fig:detector}(b). Three conically shaped polyether-ether-ketone (PEEK) spacers, placed beneath the crystals, provide support with minimal thermal losses. Four copper tabs on the top of each frame firmly press down the crystals to keep them stationary. A 400~nm thick gold film deposited on the lower surface of the crystal serves as a phonon collector, which is connected to a metallic magnetic calorimeter (MMC); this measures the rise in the crystal temperature induced by radiation absorption~\cite{gbkim2017}. A detachable photon detector is installed at the top of the copper frame. It consists of a 2-inch, 300 $\mu$m thick germanium wafer, used as a scintillation light absorber. 
A 65 $\mu$m thick Vikuiti enhanced specular reflector film (VM2000) surrounds the side and bottom surfaces of the crystal, as shown in Fig.~\ref{fig:detector}(c). 
The masses and surface areas of the Vikuiti reflectors are listed in Table.~\ref{tab:vikuiti} and the details of the AMoRE-Pilot detector system are described in Ref.~\cite{alenkov2019}.

Backgrounds from $\alpha$ decays can originate from radioactive contaminations, which are internal to and on the surface of the crystals or nearby materials directly facing them, such as the reflector and photon detector copper frame. The copper frame on the side and bottom of the crystal is located outside of the 65~$\mu$m reflector film, which renders the penetrating $\alpha$ background contributions from these parts negligible. 
However, the photon detector frame directly faces the crystal from the top, without any intervening reflector material. Both the photon detector frame and the reflector are included in the simulations. 

The event selection requirements for single-hit events (i.e., events that have signals in only one of the crystals and none in any of the other crystals) are described in the next section.

\section{Radioactive contamination of CMO crystal scintillators} 
\label{sec:3}
\begin{figure}[t]
\centering
\includegraphics[width=0.5\textwidth]{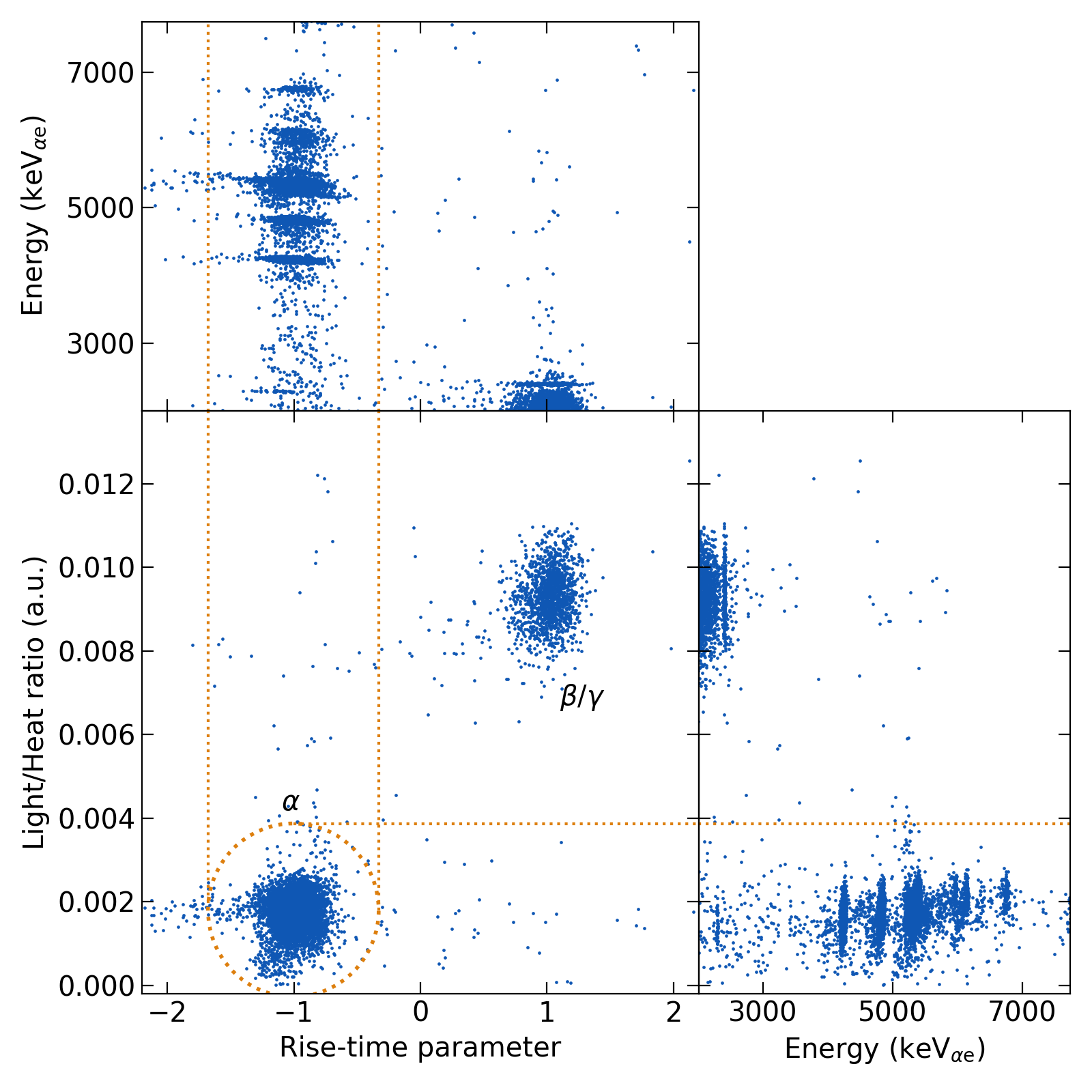}
\caption{The $\alpha$-event selection requirements based on the rise time parameter and light/heat ratio for C4.
The rise-time is normalized for $\beta/\gamma$ and $\alpha$ to be aligned at 1 and -1, respectively.
Events in the ellipse in the bottom-left are the selected $\alpha$-events.
The slope of the Light/Heat ratio on energy in the right-hand side panel is due to the energy dependence of quenching of the scintillation light output for alpha particles~\cite{Tretyak2010}.
}\label{fig:alpha_selection}
\end{figure}
The energy released in the $\alpha$-decay of an internal radio-contaminant is fully contained in the crystal and produces a peak at its Q-value~(Q$_{\alpha}$) in the energy spectrum.
The Q values of all alpha decays in the decay chains of $^{238}$U, $^{232}$Th, and $^{235}$U are higher than 4 MeV, where we expect only a small portion of beta or gamma-like events primarily generated by untagged muons or neutrons.
In this analysis, we discriminate alpha particles with energy above 2.5~MeV even though the energy degraded alphas could fall below 2.5~MeV in the continuous spectrum. This is because the limited alpha/beta discrimination at the lower energy side would increase the analysis uncertainty.
Therefore, the $\alpha$-induced-events in the 2.5--8~MeV energy range are selected using the rise time~(RT) and the light-heat ratio~(LH) distributions as shown in Fig.~\ref{fig:alpha_selection}. 
The rise time is the time between 10\% and 90\% of the signal height, which is $\sim$1--2~ms for the six crystals under study. The light-heat ratio is the ratio between photon signal and phonon signal in the event~\cite{alenkov2019,gbkim2015}. 
In these distributions, it can be seen that the $\alpha$-events are very localized at the projected LH versus RT values and there is no strong correlation between them. To safely choose $\alpha$-events with a negligible contamination fr om $\beta$/$\gamma$-events, a selection is made for each channel, which is based on the median and standard deviation for the RT and LH distributions, which are specific to that crystal detector. $\alpha$ events are accepted if they are in an ellipse that is with 10 standard deviations of the RT and LH mean values (see Fig.~\ref{fig:alpha_selection}).
\begin{figure}[t]
\centering
\includegraphics[width=0.47\textwidth]{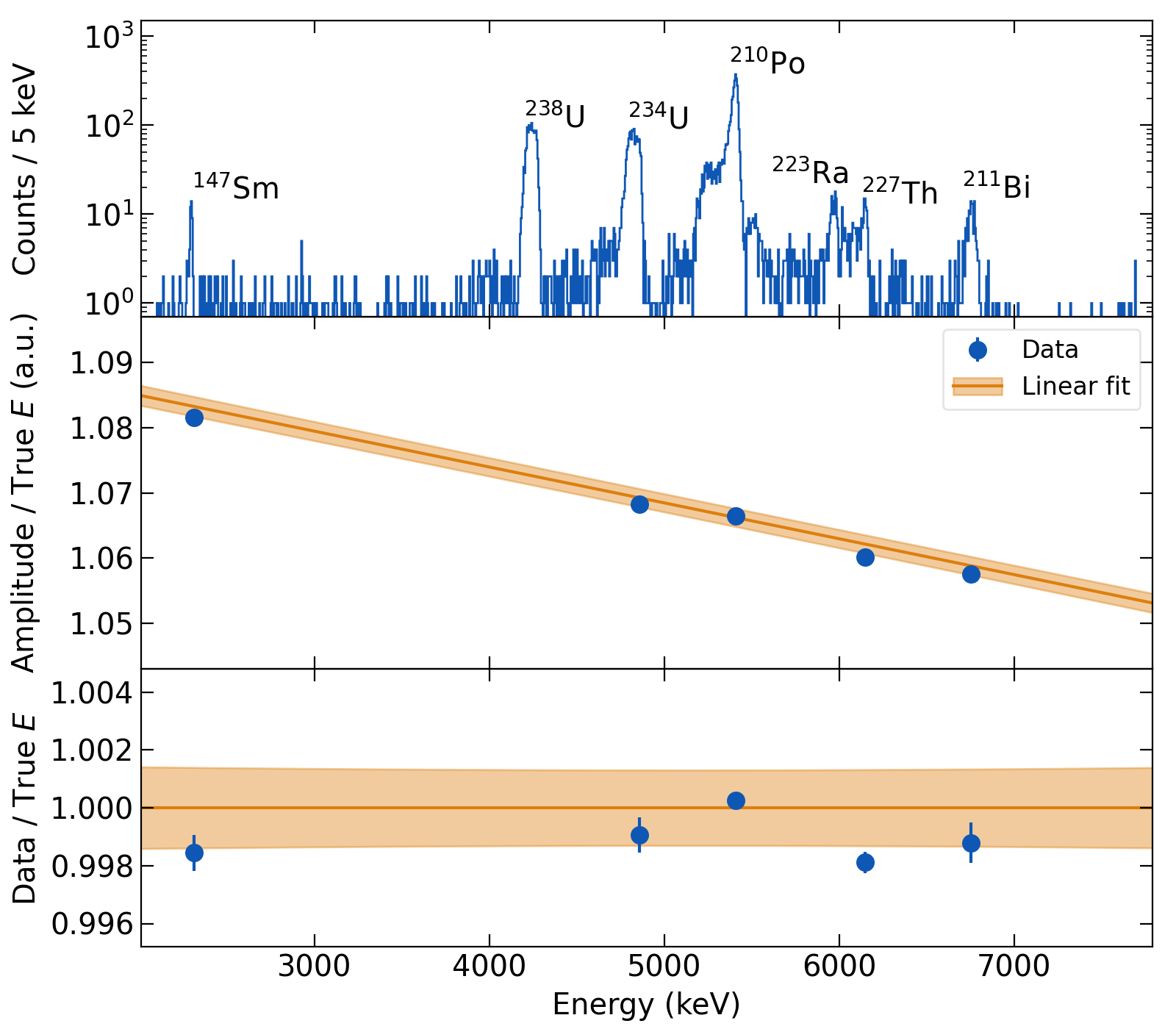}
\caption{Energy calibrations of the $\alpha$-peaks in C4. The top panel shows the energy spectrum; the middle panel shows the ratio of the true energy to the phonon signal amplitude and its linear fit; the ratio of the reconstructed energy to its nominal value is shown in bottom.}\label{fig:alpha_ecalib}
\end{figure}
\begin{figure}[t]
\centering
\includegraphics[width=0.47\textwidth]{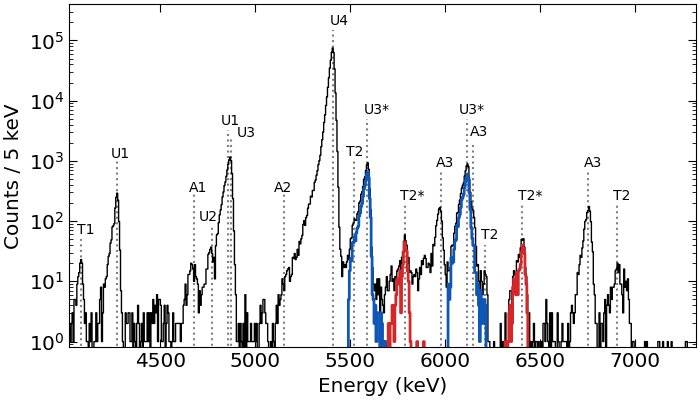}
\caption{Energy spectrum of $\alpha$-events in C2. Each dotted line and the annotation on top of it represent Q$_{\alpha}$ and the corresponding decay sub-chain. Pairs of $\alpha$-$\alpha$ coincident events in the $^{238}$U and $^{232}$Th series are annotated with stars(*) and indicated by the blue and red histograms, respectively. See Table~\ref{tab:table3} for more detail.}\label{fig:alpha_S35}
\end{figure}
\begin{table*}[t]
\begin{center}
\caption{
Radioactive contamination of the CMO crystal scintillators (see the text). 
They are obtained with a rough estimate integrating a $\pm$40 keV range around peaks.
U3* and T2* are obtained by $\alpha$-$\alpha$ coincidence events.
The numbers in the parentheses are only statistical uncertainties.}
\label{tab:table3}
\begin{tabular}{ccccccc}
\hline
Decay sub-chain & \multicolumn{6}{c}{Activity (mBq/kg)}\\
   & Crystal 1 & Crystal 2 & Crystal 3 & Crystal 4 & Crystal 5 & Crystal 6 \\ \hline
$^{238}$U--$^{230}$Th~(U1) & 0.83(2)     & 0.82(2) &   0.216(9)      & 0.51(2)      	& 0.60(2)		&1.44(2)\\
$^{230}$Th--$^{226}$Ra~(U2) & 0.18(1)   & 0.24(1)	& 1.01(2)		& 0.24(1)		& 0.47(1) 		&0.246(9)\\
$^{226}$Ra--$^{210}$Pb~(U3*) & 0.029(4) & 2.40(4)	& 0.008(2)		& 0.010(3) 	& 0.033(4)		&0.008(2)\\
$^{210}$Pb--$^{206}$Pb~(U4) & 6.51(6)   & 227.7(4)& 0.54(1)		& 1.71(4)	 	& 4.59(4)		&1.38(2)\\
$^{232}$Th--$^{228}$Ra~(T1) & 0.016(3) & 0.076(7) & 0.013(2)	& 0.009(3)		& 0.015(3) 	&0.005(1)\\ 
$^{228}$Th--$^{208}$Pb~(T2*) & 0.006(2) & 0.135(9) & 0.0003(3)	& 0.002(1)		& 0.007(2) 	&0.003(1)\\ 
$^{235}$U--$^{231}$Pa~(A1) & 0.032(4)   & 0.098(8) & 0.045(4)	& 0.037(5)		& 0.081(6) 	&0.031(3)\\ 
$^{231}$Pa--$^{227}$Ac~(A2) & 0.039(5) & 0.077(7) & 0.036(4)	& 0.027(4)		& 0.038(4) 	&0.015(2)\\ 
$^{227}$Ac--$^{207}$Pb~(A3) & 0.25(1) & 0.94(2) & 0.079(5)		& 0.059(6) 	& 0.31(1) 		&0.053(4)\\ \hline
\end{tabular} 
\end{center}
\end{table*}

Energies of the $\alpha$-events are calibrated using a linear function that fits the ratio of the phonon signal amplitudes to the true energies, which are well-known Q-values for their corresponding decay~\cite{gbkim2017}, as shown in the middle panel of Fig.~\ref{fig:alpha_ecalib} for a crystal, C4. 
The $^{210}$Po peak at 5407.5~keV was most prominent in all crystals. The peaks of $^{238}$U and $^{234}$U are detected for all crystals. However, some of them are distorted due to the contribution from surface contamination and could not be used in the energy calibration. 
Also, though three peaks above 5.7~MeV, shown in the top panel of Fig.~\ref{fig:alpha_ecalib}, were attributed to the decays of the $^{227}$Th chain, 
the $^{223}$Ra peak was not used in the calibration because it became an outlier of the calibration curve due to the surface contribution around the peak that made the peak fit incorrect.
The peak at 2.3~MeV, which is present in the spectra of all detectors was identified as $^{147}$Sm decay events. At least four alpha peaks were used for all crystals to obtain the calibration function. The peak amplitudes were determined by fitting them to a Gaussian distribution at a narrow amplitude window around the peak to avoid the contributions from surface contaminations.  
The uncertainty in this calibration was obtained by quadratic summing the standard deviation of calibration points from the fit curve and a linear fit error. As shown in the bottom plot of Fig.~\ref{fig:alpha_ecalib}, the difference between the fit function and the actual data points is about 0.1\% on average. It is considered a systematic error for the energy scale in the background modeling.

After applying the energy calibration as discussed, the energy resolutions of known $\alpha$ peaks were obtained by a fit with an exponentially modified Gaussian function~\cite{alenkov2019}.

Figure~\ref{fig:alpha_S35} shows the energy spectrum of the $\alpha$-events in C2. 
There are $\alpha$ peaks from the decay of $^{235}$U in the crystals with the activities similar to the $^{238}$U or the $^{232}$Th chains. The natural abundance of $^{235}$U, 0.7\%, predicts about 4.6\% of $^{235}$U activity compared to $^{238}$U. Therefore, this high activity of $^{235}$U in these crystals is exceptional. 
It is not clearly understood how the contamination occurred.\footnote{The anomalous activity of the $^{235}$U chain may be explained by effect of the enrichment processes of molybdenum and calcium or by the proximity of the stable isotopes enrichment plant to a $^{235}$U enrichment facility}

The internal background activities are initially estimated by counting the number of events within a $\pm$40~keV energy window for each corresponding Q$_{\alpha}$. 
However, individual alpha peaks from the decays of $^{235}$U, $^{238}$U, and $^{232}$Th partially overlap each other in the spectrum and the contamination of the crystals by $^{235}$U is rather high. In most of the alpha active, nuclides of the $^{235}$U chain alpha decays go with a significant probability to the excited levels of the daughters with emission of gamma quanta (and conversion electrons). 
In addition, there are contaminations at the surfaces of the crystals and Vikuiti films.
Therefore, $\alpha$-peak counting rates obtained in this way are overestimated and we tried to get the bulk contamination level for each crystal by studying alpha-alpha time-correlated events~\cite{Azzolini2021}.   

For the decay sub-chains where one or more isotopes have lifetimes on the scale of minutes, such as $^{218}$Po~($T_{1/2}=3.10$~min) and $^{220}$Rn~($T_{1/2}=55.6$~sec), sequences of $\alpha$-decays can be identified. 
A coincident event can be reconstructed by finding an $\alpha$ event followed by a delayed coincidence $\alpha$ event in the 5~$T_{1/2}$ time window with the energy constraint for each Q$_{\alpha} \pm$~40~keV.
The activity levels of these sub-chains, such as U3* and T2*, can be more precisely determined by analyses of these coincidence events. If the second alpha occurs less than 100~ms from the first alpha, it would be excluded since the energy of the second alphas would not be accurately estimated. The rejection probabilities of the second alphas are estimated as 0.04\% and 0.1\% for Q$_\alpha$ = 6.115 and 6.405 MeV respectively, which are negligible. The energy window will include all the bulk contamination and a part of deep surface contamination, which will be discussed in the background modeling section. 

A summary of the radioactive contamination for the CMO crystals obtained by estimates of the corresponding $\alpha$-peak counting rates is given in Table~\ref{tab:table3}; the active-$\alpha$ nuclides from the sub-chains, U3* and T2*, were measured by $\alpha$-$\alpha$ coincidence events.
A more quantitative analysis that includes surface contamination is described in Sect.~\ref{sec:5}.

\section{Background simulations}
\label{sec:4}
\sloppy
Monte Carlo~(MC) Simulations have been conducted using Geant4~\cite{geant4} version 10.4.2, in which we have implemented an AMoRE-specific physics list for both internal and surface background simulations. We adopted the G4ScreenedNuclearRecoil class~\cite{screenednuclearrecoil} for the simulation of low-energy nuclear recoils from $\alpha$ surface contaminations, the G4EmLivermorePhysics class for low energy electromagnetic processes, and the G4RadioactiveDecay for radioactive decay processes. 
 
To model the measured $\alpha$ energy spectrum, we simulated internal radioactive sources such as the full decay chains of $^{238}$U, $^{232}$Th, and $^{235}$U. 
The measured activities, listed in Table~\ref{tab:table3}, are used to normalize the simulation results.
In the simulation framework, each simulated event includes an energy deposit in the crystals within a 5-ms event window considering the 1$\sim$2~ms rise times for six crystals. Pile-up events from decays with short half-lives, followed by the subsequent daughter decay in the time window of 100~ms, which is a few times the typical pulse width ($\sim$20--30~ms) in cryogenic measurements~\cite{HJLee2015}, are also simulated. 
 If the second alpha occurs less than 100 ms from the first alpha, the second alpha was excluded from the MC data as it is in the experimental data.
The simulated spectrum was smeared by the energy-dependent energy resolution obtained during the calibration described in Sect.~\ref{sec:3}. 

Surface $\alpha$ contaminations of the crystal scintillators, Vikuiti films, and of the photon frames (that has partially crystal-facing parts) were simulated in order to investigate their contributions to the background.

\begin{figure}[ht]
\begin{center}
\includegraphics[width=0.48\textwidth]{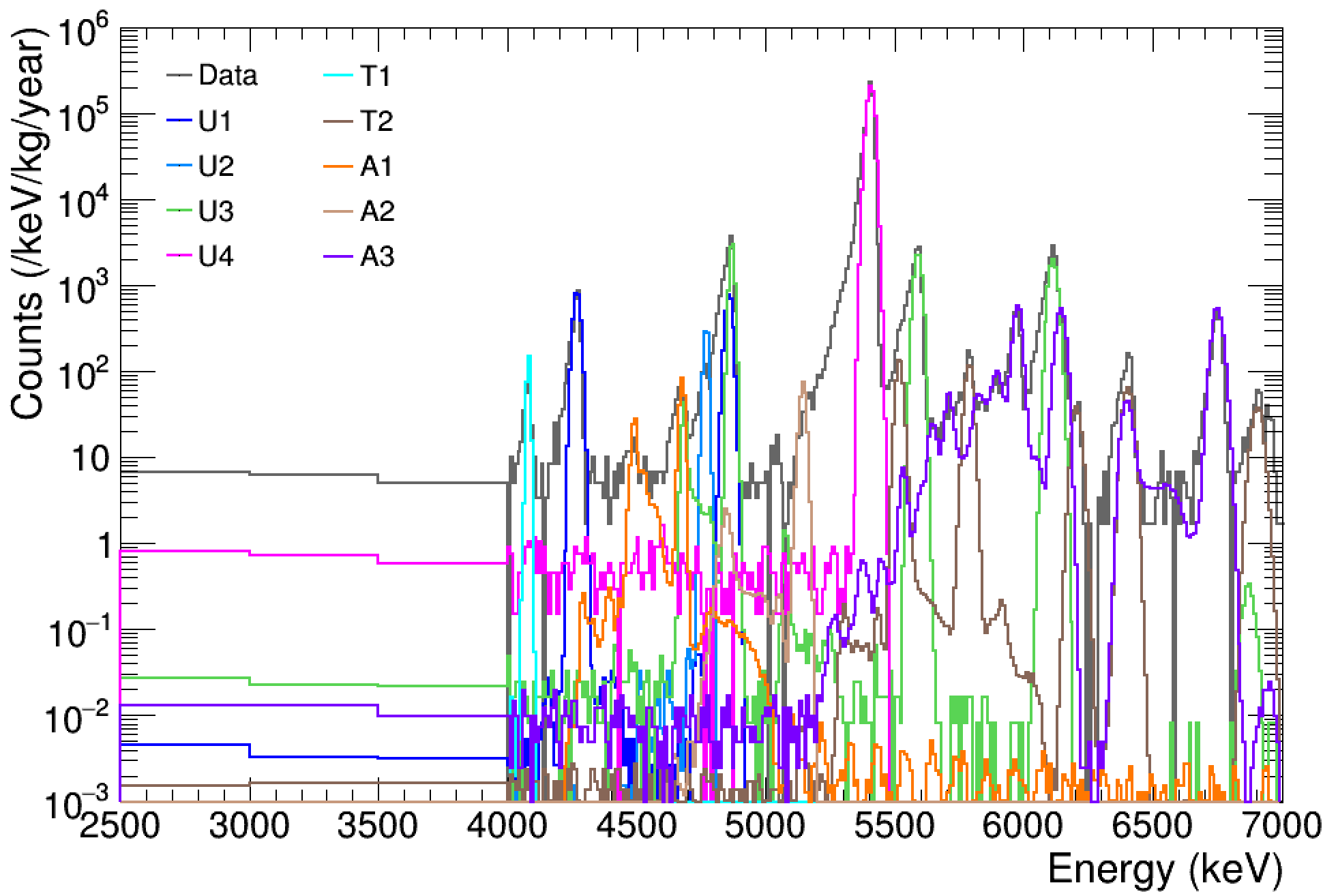}
\caption{Energy spectra of $\alpha$ events in C2. The black solid line represents the measured $\alpha$ energy distribution and other colors represents the simulation results for internal radionuclides.
}
\label{fig:internal}
\end{center}
\end{figure}
\begin{figure}[ht]
\begin{center}
\includegraphics[width=0.43\textwidth]{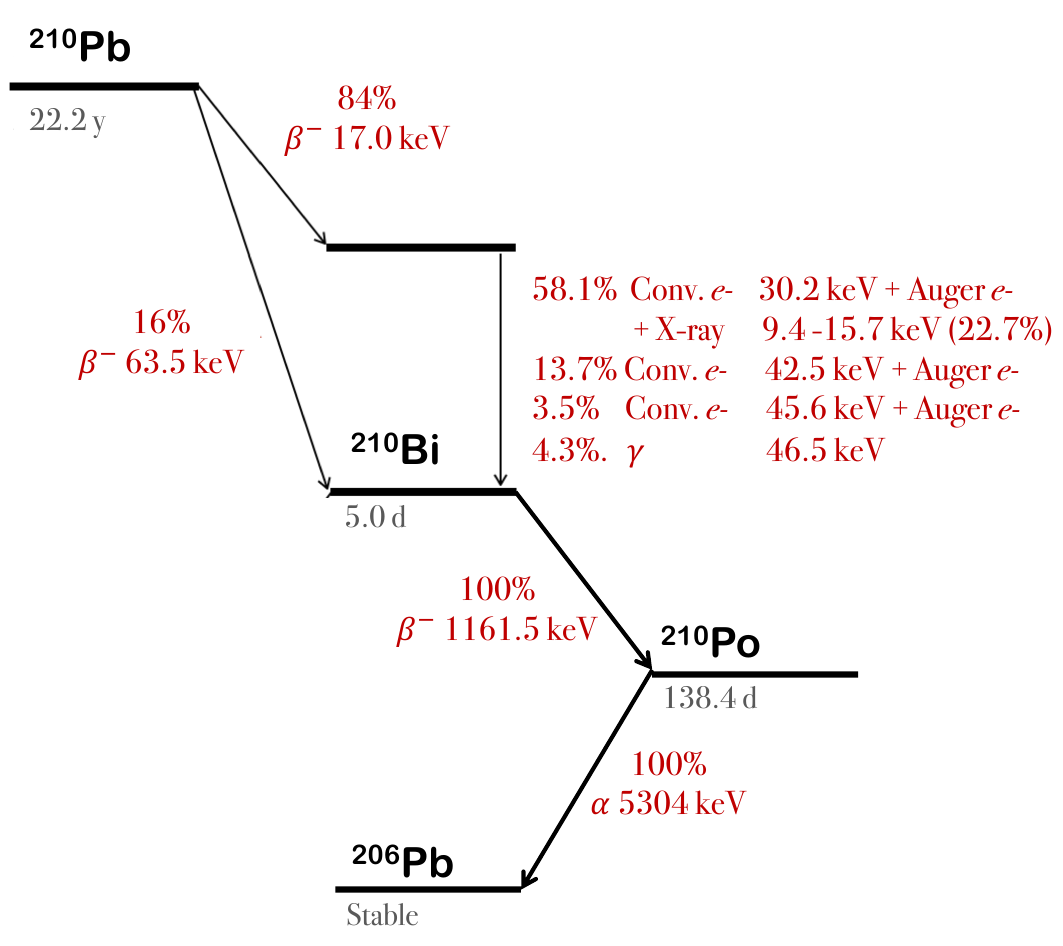}
\caption{
$^{210}$Pb decay chain.
The scheme is built by using data from the Brookhaven National Nuclear Data Center (https://www.nndc. bnl.gov/) 
}
\label{fig:pb210}
\end{center}
\end{figure}
\begin{figure*}[t]
\begin{center}
\begin{tabular}{cc}
\includegraphics[width=0.45\textwidth, height=0.2\textheight]{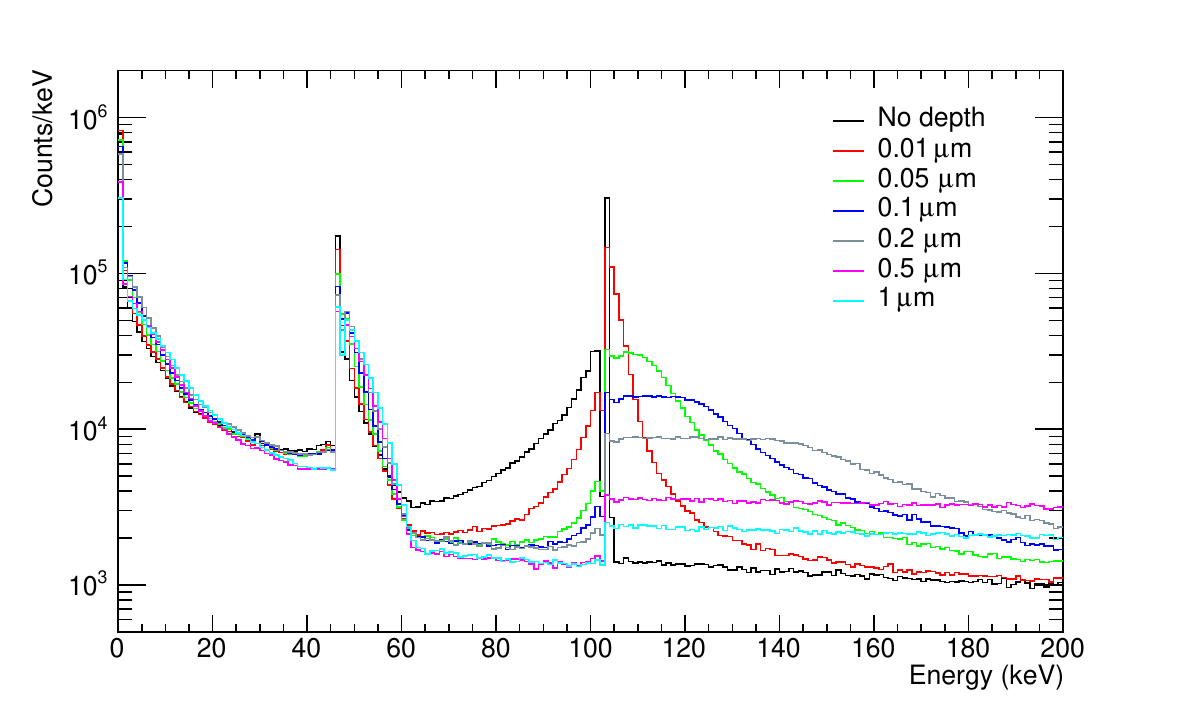} &
\includegraphics[width=0.45\textwidth, height=0.2\textheight]{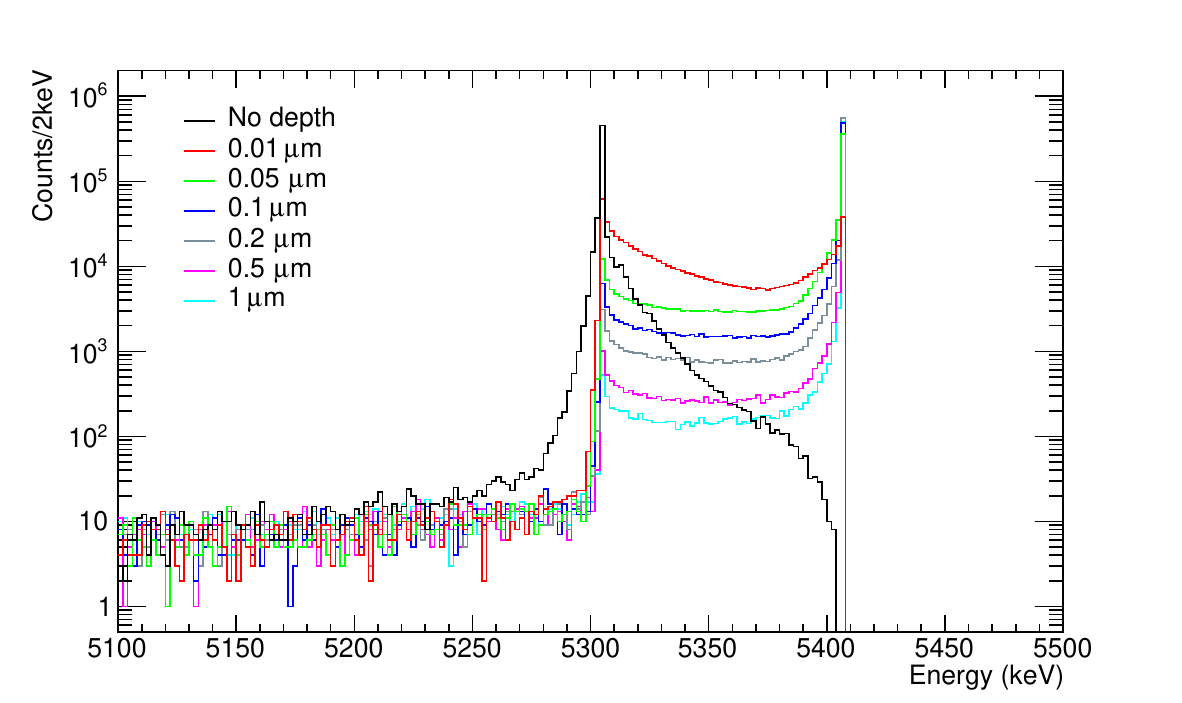} \\
(a) & (b)  \\
\includegraphics[width=0.45\textwidth, height=0.2\textheight]{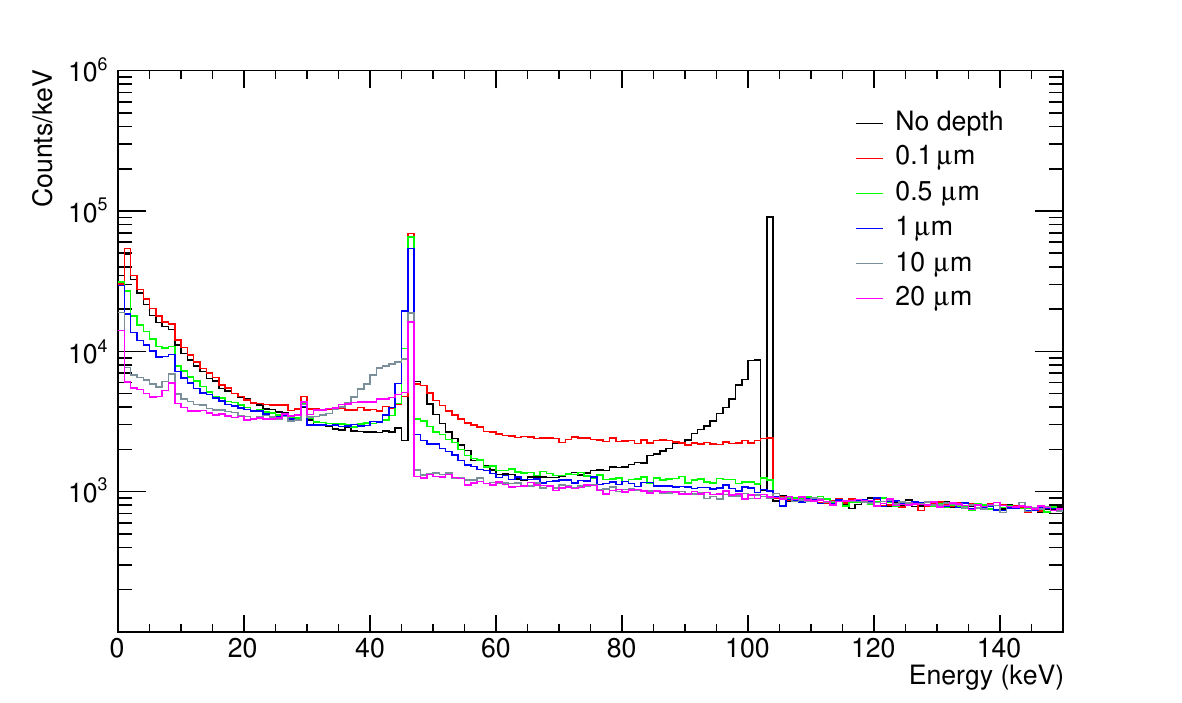}&
\includegraphics[width=0.45\textwidth, height=0.2\textheight]{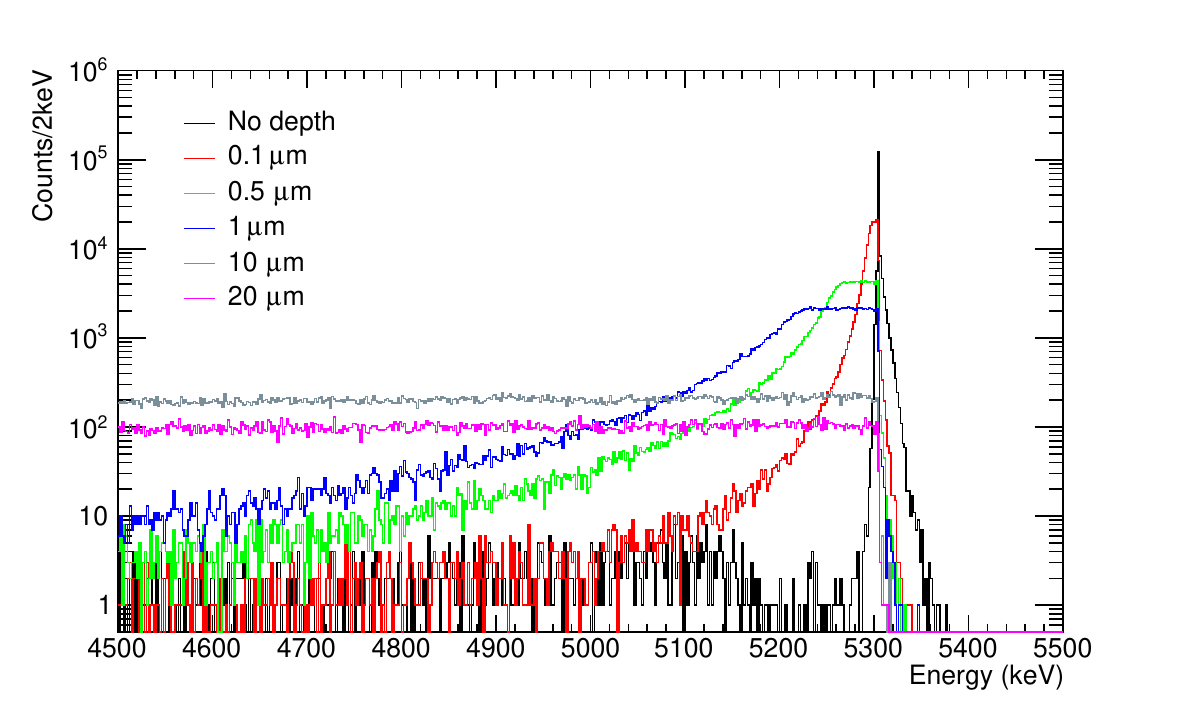}\\
(c) & (d) \\
\end{tabular}
\caption{ 
(a) Low- and (b) high-energy spectra from surface $^{210}$Pb contaminants distributed uniformly within various surface depths of a crystal. The Q-value peak at 5407.5~keV that is the sum of 5304.3~keV $\alpha$ and 103.2~keV $^{206}$Pb recoil is affected by energy loss depending on the crystal surface depth. 
(c) Low- and (d) high-energy spectra from surface $^{210}$Pb contaminants distributed uniformly within various surface depths of Vikuiti film. An $\alpha$ from the Vikuiti surface leaves an amount of energy in the crystal that can be as high as 5304~keV. Some spectral features like plateaux appear from surface contaminants within the depth on the scale of or less than 1~$\mu$s.
}
\label{fig:surfacesimulation}
\end{center}
\end{figure*}

\subsection{$\alpha$ energy spectra}
\label{sec:4.1}
Internal radionuclides from the decay chains of $^{238}$U, $^{232}$Th, and $^{235}$U were simulated and their event rates were normalized by using the measured activities. Figure~\ref{fig:internal} shows the comparison of the measured $\alpha$ spectrum~(black) to spectra~(other colors) simulated in the energy region of (2.5--7)~MeV.
In the simulation, we assumed that radionuclides in a sub-chain are in equilibrium. 
As listed in Table~\ref{tab:table3}, 
there are four sub-chains for $^{238}$U decays~(U1--U4), two sub-chains for $^{232}$Th decays~(T1--T2), and three sub-chains for $^{235}$U decays~(A1--A3) shown in different colors, respectively.
All peaks with the same color are from one sub-chain with the same activity.
As shown in Fig.~\ref{fig:internal}, the overall measured spectrum is not fully described by internal radionuclides alone, even though the typical $\alpha$ peak positions are well matched. In particular, the level of the continuum background from 2.5~MeV to 4~MeV in the measured energy spectrum is higher than the simulated spectra. This implies that there are additional contributions from surface $\alpha$ contaminations on the crystals and the nearby materials. 
Moreover, the continuum distribution lower than the 5407.5~keV peak of $^{210}$Po indicates a substantial contribution from $^{210}$Po surface contaminations.
In the following section, the details of the spectrum changes as a function of the contaminant depth are described.

\subsection{Surface $\alpha$ contaminations}
\label{sec:4.2}
Figure~\ref{fig:pb210} shows the decay scheme of the $^{210}$Pb decay chains. It decays to $^{210}$Bi, which decays to $^{210}$Po and subsequently contributes to the high-energy spectra by the $\alpha$ decay to $^{206}$Pb.
If $^{210}$Pb and $^{210}$Po are embedded in the surface they could act as sources of low-energy background because of their long half-lives of 22.3~years and 138~days, respectively.
Because the beta decay to $^{210}$Bi results in low-energy events via electrons and $\gamma$/X-ray emissions, the spectral features of these events for energies less than 60~keV depend on their embedded depth.
$^{210}$Po decays via the emission of 5304.3~keV $\alpha$ particle and a $^{206}$Pb recoil nucleus with energy 103.2~keV.
If the $\alpha$ particle from the decay of $^{210}$Po escapes the surface without energy deposition, the recoiling $^{206}$Pb creates a low energy signal that depends upon its depth. Alternatively, the 5304~keV $\alpha$ energy can be deposited in the crystal surface when the $^{206}$Pb surface recoil escapes. 
It is also possible that both alpha and recoil deposit all the energy in the crystal and make an event at Q-value.

To understand the background contributions from surface $\alpha$ particles as a function of their contamination depth, we simulated $^{210}$Pb by generating it at random locations within various surface depths for a crystal and its Vikuiti film. The low- and high-energy spectra from the decay of $^{210}$Pb at both crystal and Vikuiti surfaces for various surface thicknesses are shown in Fig~\ref{fig:surfacesimulation}(a), (b), (c), and (d), respectively.
Because the path length of the $^{206}$Pb, which recoils with 103~keV kinetic energy, can be as small as $\sim$50~nm, calculated with SRIM~\cite{SRIM}, inside the crystal, the Q-value peak at 5407.5~keV, which is the sum of 5304.3~keV $\alpha$ and 103.2~keV $^{206}$Pb recoil, is affected by energy loss depending on the crystal surface depth, as shown in Fig~\ref{fig:surfacesimulation}(a) and (b).
For very shallow $\alpha$ decays, it shows up at 5304~keV as a peak, while the 103 keV peak shows up at low energies.
When decayed alphas with no depth move into the crystal, the recoiled nucleus moves towards the Vikuiti reflector. Then, though it is rare, the nucleus is reflected in the crystal and deposits energy smaller than the recoil energy, so the energy deposition will be higher than the alpha energy as shown in Fig~\ref{fig:surfacesimulation}(a).
The range of a 5~MeV $\alpha$ particle inside a CMO crystal is $\sim$25~$\mu$m while that inside the Vikuiti film is $\sim$40~$\mu$m.
For $\alpha$ decays at deeper positions, the spectra becomes similar to internal $\alpha$ decays.  
There are also low-energy contributions around 50~keV due to the beta decay of $^{210}$Pb.
These are attributed to 46.5~keV $\gamma$ emissions together with energies up to 17~keV, emitted in the transition to the excited state of $^{210}$Bi.
As shown in Fig.~\ref{fig:surfacesimulation}(d), an $\alpha$ from the Vikuiti surface leaves an amount of energy in the crystal that can be as high as 5304~keV. If the $\alpha$ event escapes from the depth deeper than about 10~$\mu$m it contributes to the flat backgrounds with no peak near 5.3~MeV, while some spectral features like plateaux appear from surface contaminants uniformly distributed within the depth on the scale of or less than 1~$\mu$m. Thus, we used both shallow and deep depth surface simulations for $^{210}$Pb on Vikuiti surfaces in the $\alpha$ background modeling. 

\section {Modeling $\alpha$ spectra}
\label{sec:5}	
\subsection{$\alpha$ spectra of internal and surface contamination}
\label{sec:5.1}
\begin{figure*}[ht]
\begin{center}
\includegraphics[width=0.9\textwidth]{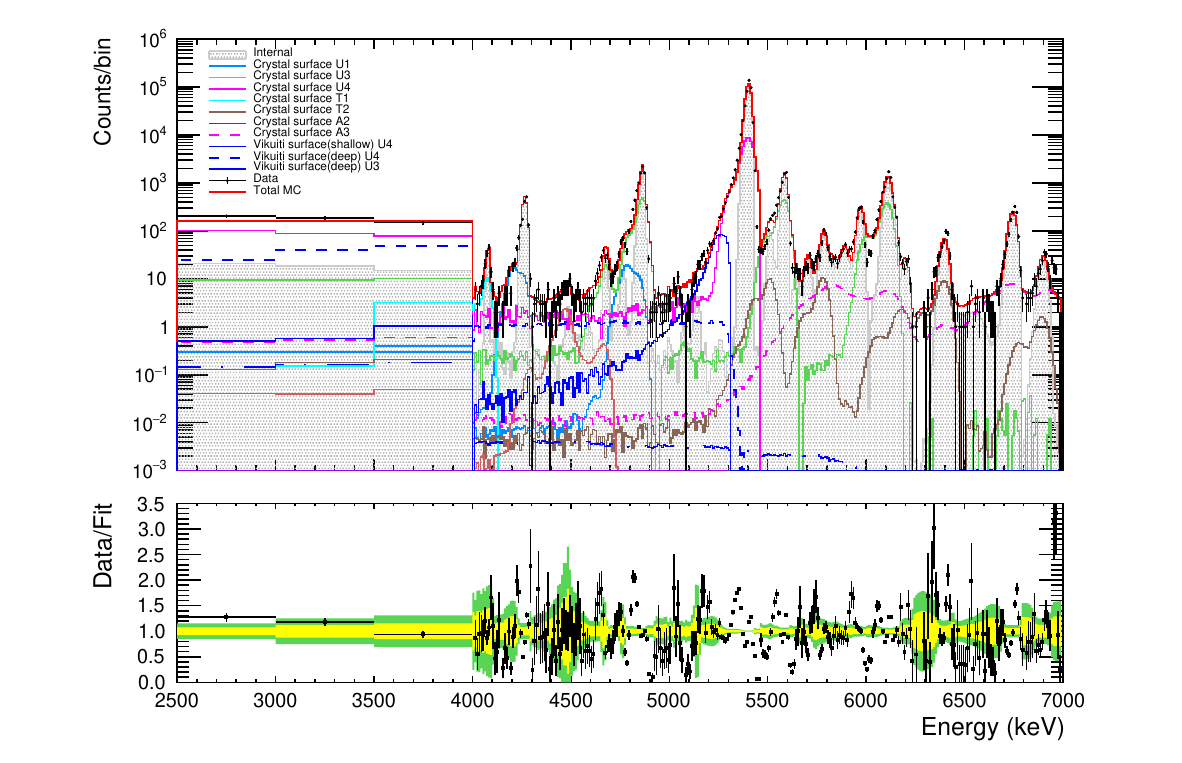}
\caption{Energy spectra of single-hit $\alpha$ events in C2. The simulation results are fitted to the measured data (top panel).
The lower panel shows the bin by bin ratios of data to fitted result; the colored bands centered at 1 in the lower panel represent 1 and 2$\sigma$ of the total uncertainty obtained by the quadratic summing systematic uncertainty from the energy scale and uncertainty of fit parameters from the modeling. The uncertainty of the data/fit ratio is the statistical error. 
}
\label{fig:fitresult}
\end{center}
\end{figure*}
\begin{figure}[ht]
\begin{center}
\hspace*{-1cm}\includegraphics[width=0.47\textwidth]{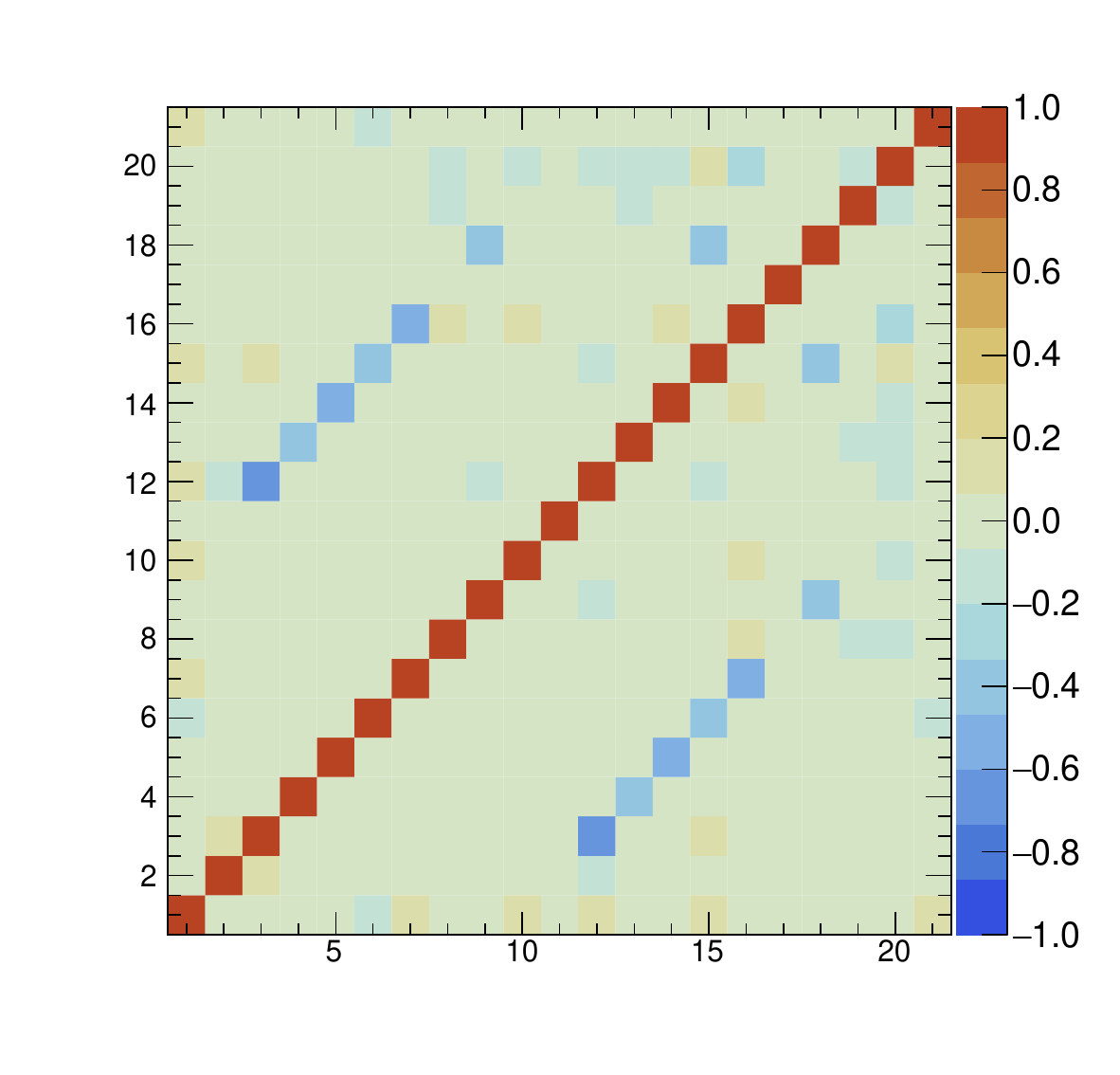}
\caption{Correlation matrix among fitted activities in C2.}
\label{fig:matrix}
\end{center}
\end{figure}
\begin{table*}[!b]
\begin{center}
\caption{
Fitted activities of $^{232}$Th, $^{235}$U and $^{238}$U sub-chains in the CMO crystal scintillators.
}
\label{tab:tabelfitbulk}
\begin{tabular}{cccccccc}
\hline
Decay sub-chain & No. & \multicolumn{6}{c}{Activity (mBq/kg)}\\
      &  & Crystal 1 & Crystal 2 & Crystal 3 & Crystal 4 & Crystal 5 & Crystal 6 \\ \hline      
      $^{238}$U--$^{230}$Th~(U1)    &1  & 0.57(4)    & 0.77(6)   & 0.015(11) & 0.13(3)      	& 0.18(3)		&0.64(5) \\
      $^{230}$Th--$^{226}$Ra~(U2)  &2 & $<$0.003  & 0.02(2) & $<$0.01 & $<$0.003	         &  $<$0.002	&$<$0.002 \\
      $^{226}$Ra--$^{210}$Pb~(U3*) &3 &  $<$0.002  & 1.94(8)   & 0.004(6) &  $<$0.007 	& $<$0.005	&$<$0.005\\
      $^{210}$Pb--$^{206}$Pb~(U4)  &4 & 5.1(2) & 186.1(1.1) & 0.13(6)	& 0.9(1)	 	& 2.6	(2)		&0.37(12) \\
      $^{232}$Th--$^{228}$Ra~(T1)  &5 & 0.001(13)  & 0.03(2) & $<$0.01 &  $<$0.006	& $<$0.01 	&$<$0.004\\ 
      $^{228}$Th--$^{208}$Pb~(T2*)  &6 & $<$0.005 & 0.14(2) & $<$0.0003 & $<$0.001	& $<$0.004 	&$<$0.002\\ 
      $^{235}$U--$^{231}$Pa~(A1)   &7 & 0.017(13)  & 0.01(2) & 0.001(30)	& 0.01(2)		& 0.01(2)		&0.005(9)\\ 
      $^{231}$Pa--$^{227}$Ac~(A2)  &8 & $<$0.006 & 0.01(1) & $<$0.003	& $<$0.006	& 0.01(1) 		&0.002(11)\\ 
      $^{227}$Ac--$^{207}$Pb~(A3)  & 9 & 0.37(3)  & 0.65(4) & 0.09(1)	& 0.054(14)	& 0.43(3)		&0.061(12) \\ \hline
\end{tabular}
\end{center} 
\end{table*}
\begin{table*}[ht]
\begin{center}
\caption{
Fitted activities of $^{232}$Th, $^{235}$U, and $^{238}$U sub-chains in the surface layers of CMO crystal scintillators and reflecting film ($\mu$Bq/cm$^{2}$).
Surface depths are in units of $\mu$m.}
\label{tab:tabelfitsurface}
\setlength{\tabcolsep}{1pt}
\begin{tabular}{c c c c c c c c c c c c c c c}
\hline
Source & Decay & No. & \multicolumn{2}{c}{Crystal 1} & \multicolumn{2}{c}{Crystal 2} & \multicolumn{2}{c}{Crystal 3} & \multicolumn{2}{c}{Crystal 4} & \multicolumn{2}{c}{Crystal 5} & \multicolumn{2}{c}{Crystal 6}\\
  &  sub-chain &  & Depth & Activity & Depth & Activity & Depth & Activity & Depth & Activity & Depth & Activity& Depth & Activity\\
\hline
Crystal & $^{238}$U--$^{230}$Th~(U1)      & 10  & 0.01   & 1.79(27) &0.01& 0.52(18) 	& 0.002 &8.8(4) 	& 0.01  & 3.5(4)     	& 0.01    	& 8.7(6) 		& 0.01 & 9.8(7)\\
  	    &  $^{230}$Th--$^{226}$Ra~(U2)  & 11 & -         & -              &  -     & -           		& 0.002 & $<$0.4 	&0.002 & 0.1(1)	     	& 0.002 	& 0.68(29) 	& - & -\\
	    &  $^{226}$Ra--$^{210}$Pb~(U3*) & 12& 0.01  & 0.47(12)   & 0.01& 11.7(7) 		& 0.005 & 0.7(2)	&0.002 & 0.67(15)  	& 0.002  	& 3.6(4) 		& 0.002 & 0.07(2) \\
  	    &  $^{210}$Pb--$^{206}$Pb~(U4)  & 13 & 0.1    & 4.57(46)   & 0.1 & 108.4(2.6) 	&0.1 & 3.2(5)  		& 0.03  & 2.83(37)  	& 0.1		& 14.7(1.3) 	& 0.08 & 6.9(1.0)\\
  	    &  $^{232}$Th--$^{228}$Ra~(T1)  & 14 & 0.05  & 0.05(5)  & 0.05 & 0.2(1) & 0.05  & 	$<$0.1		& 0.01  &  0.03(5)  	& 0.01	&0.1(1) 		& 0.01 & 0.07(7)\\
  	    &  $^{228}$Th--$^{208}$Pb~(T2*) &15 & 0.05  & 0.23(7) & 0.05 & 0.31(15)   		&0.05 &  $<$0.4 	& 0.002 &0.15(8)   	& 0.005	&0.98(27) 		& 0.04 &0.51(17)\\
  	    &  $^{235}$U--$^{231}$Pa~(A1)    & 16 & 0.01 & 0.1(1)   & 0.01      & 0.1(2)		&0.01 & 0.82(33) 	& 0.01   & 0.2(2)    	& 0.05	&0.68(37) 		& 0.05 & 1.04(27)\\
  	    &  $^{231}$Pa--$^{227}$Ac~(A2)  & 17 & 0.05  & 0.26(4) & 0.05& $<$ 0.08&	 0.05 	& 0.33(4)		& 0.05   & 0.34(5)  	& 0.05 	&0.1(1)		& 0.015 &0.09(9)\\
  	    &  $^{227}$Ac--$^{207}$Pb~(A3)  & 18 & 0.05  & 0.2(1) & 0.05 & 0.52(15) 		& 0.05 & 0.46(12) 	& 0.05   & 0.32(6)  	& 0.05	&1.1(2) 		& 0.1 & 0.2(1)\\
Vikuiti  & $^{210}$Pb--$^{206}$Pb~(U4)   & 19 & 0.8  & 9.6(9) & 0.8  & 6.7(1.2)	& 0.9 	& 3.98(51) 		& 0.9    & 10.5(1.6) 	& 0.7		&0.5(1) 		& 0.1 &7.9(8)\\
 film  	& $^{210}$Pb--$^{206}$Pb~(U4) & 20 & 20   & 4.2(9) & 20 & 2.9(1.4) 		&20 & 11.2(1.5) 	& 20     & 2.6(7)      	& 20		&8.4(1.3) 		& 20 &1.29(69)\\
	         & $^{226}$Ra--$^{210}$Pb~(U3) & 21 & 20   & 0.004(4) & 20 & 0.004(3) 		&20 & 0.005(2)		& 20     & 0.004(2)  	& 20		&0.004(2) 		& 20 &0.004(3)\\ \hline  
\end{tabular} 
\end{center} 
\end{table*}
\begin{figure}[t]
\begin{center}
\hspace*{-0.2cm}\makebox[0.5\textwidth]{\includegraphics[width=0.5\paperwidth]{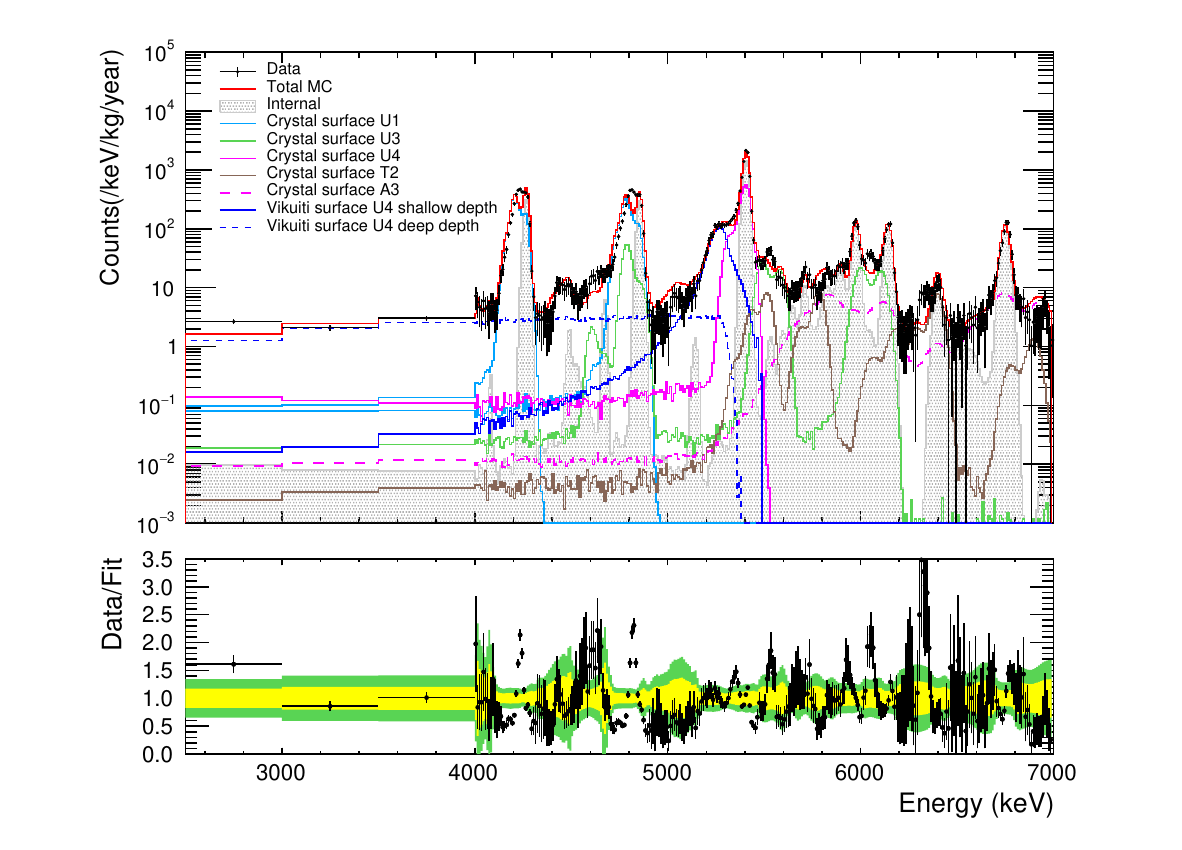}}
\caption{Averaged energy spectrum of single-hit $\alpha$ events for the five crystal detectors~(C2 is excluded). 
The simulation results are compared with measurements of the averaged background rates weighted by crystal mass, effective live time of the detector, and dead-time correction (top panel).
The lower panel shows the bin by bin ratios of data to fitted result; the colored bands centered at 1 in the lower panel represent 1 and 2$\sigma$ of the total uncertainty obtained by the quadratic summing systematic uncertainty from the energy scale and uncertainty of fit parameters from the modeling. The uncertainty of the data/fit ratio is the statistical error. 
}
\label{fig:fitresultall}
\end{center}
\end{figure}
 We first simulated background spectra of single hit $\alpha$ particles from internal and surface contamination in nine sub-chains, as follows. Four sub-chains for $^{238}$U decays, two sub-chains for $^{232}$Th decays, and three sub-chains for $^{235}$U decays, where we assumed equilibrium in radioactivities of each sub-chain. 
Then we fitted the simulated spectra to the measured data using a binned log-likelihood method with the following formula~\cite{pdg},
\begin{align}
	 -2\ln\lambda(\theta) &=
		\begin{aligned}[t]
		&2\sum_{i=1}^{N_\mathrm{bin}}\left[\sum_{j=1}^{N_\mathrm{bkg}}\theta_jB_{ij}-n_i+n_i\ln\frac{n_i}{\sum_{j=1}^{N_\mathrm{bkg}}\theta_jB_{ij}}\right] \\		
		&+\sum_{j=1}^{N_\mathrm{bkg}}\left(\frac{\theta_j-m_j}{\sigma_j}\right)^2,
		\end{aligned}
\end{align}
where $\lambda(\theta)$ is the likelihood ratio in terms of the rates of the MC components $\theta$~=~($\theta_1$, $\theta_2$, $\cdots$, $\theta_{N_\mathrm{bkg}})$, $n_i$ is the number of events in the $i$th energy bin of the data histogram and $B_{ij}$ is the number of events in the $i$th bin of the $j$th simulation component. The last term denotes a penalty for the rate $\theta_j$ of the $j$th component and is only active if there is an independent measurement of this component; $m_j$ and $\sigma_j$ are the measured value and the error, respectively. 

The fitting range is 2.5~MeV--7.0~MeV, and the activities of each sub-chain are set as floating and/or constrained parameters in the fit. 
Conservative upper limits, which are set starting from the integration of peak counts in the experimental spectrum, are used for internal radionuclides. 

For the background contributions from surface contaminations, we simulated $^{210}$Pb embedded in the crystal surfaces and the Vikuiti film.
In addition, we considered surface contributions of all other radionuclides from the nine $^{238}$U, $^{232}$Th, and $^{235}$U decay sub-chains. 
There are peaks at alpha energy because the Q-value peaks are affected by energy loss due to nuclear recoil escapes from the shallow depth,
as described in Sect~\ref{sec:4.2}, and thus we simulated background spectra at various surface depths from 0.002 to 1~$\mu$m by generating the radioactive contaminants uniformly within the depth. 

The spectral shape of the $\alpha$-peak is not sufficiently simple to be described as allowing for two or three representative depths for six crystals.
Therefore, we performed the fit with different depths for each crystal and selected the most effective depth in each crystal in terms of the quality of the fit. 
In the final fit, we treated them as free-floating parameters to estimate their contribution to the overall background.

For the shallow depth of the Vikuiti reflector, the only spectral feature shown in the $\alpha$-spectrum is from the $^{210} $Pb decay, and it is included in the modeling fit. We also simulated the background spectrum from $^{210}$Pb distributed within a 20~$\mu$m thickness for the deep surface of the Vikuiti reflector, which contributes to the continuum in the energy spectrum and is included in the fit. 
There could be continuum parts contributed by $^{238}$U and $^{232}$Th contaminations in the Vikuiti surface of 20~$\mu$m~\cite{Azzolini19_background}.
The Vikuiti reflector was measured using the ICP-MS (Inductively Coupled Plasma Mass-Spectroscopy) method~\cite{Sala2016} and resulted in (19.8$\pm$12.4)$~\mu$Bq/kg for $^{238}$U and (11.5$\pm$5.3)~$\mu$Bq/kg for $^{232}$Th. 
We included $^{226}$Ra decay (sub-chain U3) in the fit. 

Because the influence of surface contamination distorts the spectral shape of the $\alpha$-peak, it is not easy to accurately obtain the signal amplitude corresponding to the Q value. Energy calibration using these peaks produces uncertainties on the energy scale and energy resolution. For a more quantitative statistical analysis, $\pm$1$\sigma$ of energy resolution was considered in the modeling fit as systematic uncertainties. Among other things, the energy scale is set based on the linear fit of calibration data points. It has errors propagated from the uncertainty of the calibration. Thus, we consider a coefficient in the MC spectrum for the energy scale errors~\cite{cosine2021_background}. The coefficient is determined by fitting the MC spectrum to the data and is approximately 0.14\% of the energy. 
In addition, when the bin size was changed within the range that could at least distinguish the spectral shape of the peak, the effect of these bin-size changes on the fit results is negligible.

Figure~\ref{fig:fitresult} shows fitted results for all the simulated background spectra plotted as various lines with different colors and styles for single-hit $\alpha$ events in C2. The overall energy spectrum summed over all simulations (red line) describes the data (black line), including the continuum part in the energy spectrum that extends down to the ROI region. 
The grey area represents the internal contribution and the white area below the red line represents the surface contribution, which is composed of different-color\ background compositions. 
A major contributor to the white area is surface contaminations of $^{210}$Pb on both the crystal and Vikuiti film.
The simulated spectral shapes of the internal radionuclides, as shown in the grey area, produce Gaussian peaks, whereas most of peaks in the measured data are non-Gaussian.
Including the background contributions from crystal surface contaminations improves not only the description of the non-Gaussian peaks but also the continuum between the peaks, as shown in Fig.~\ref{fig:fitresult}.
There is a region above 6 MeV where the data/model matching is worse because some of the data fluctuate due to low statistics. 
However, there are still tens of keV differences in the peak values at several energies between the data and the model, which makes the wavy shape of the residuals, even though we considered the energy scale as a systematic uncertainty in the fit. It is more discussed in Sect.~\ref{sec:5.2}.

Figure~\ref{fig:matrix} shows the correlation matrix among fitted activities in C2. The fit parameters are numbered from 1 to 9 representing internal sources and 10 to 21 surface sources. 
The bulk and surface sources representing the same crystal contaminants are somewhat highly reversely correlated. However, other crystals with different depths show small correlations except for bulk/surface $^{210}$Pb. This anti-correlation between bulk and surface sources is due to the contribution of surface contaminants to the Q-value peaks. In contrast, the different surface contaminants are either less or not correlated with each other in the six crystals.

Each of the six crystals was individually fitted to obtain six fitted energy spectra. Each spectrum was weighted by crystal mass, effective live time of the detector, and dead-time correction.
Figure~\ref{fig:fitresultall}  shows the weighted average of measured (black) and fitted (red) background spectra for five detectors except for C2.
C2 was excluded because of its extraordinary contamination of $^{210}$Pb, as shown in Fig.~\ref{fig:fitresult}.
The grey area represents the internal contributions and other colors represent the various surface contributions.
The dominant backgrounds in the continuum between 2.5~MeV and 4~MeV are due to deeply embedded surface $^{210}$Pb on the Vikuiti film, while C2 is due to $^{210}$Pb on the crystal surface. 
The continuum between the peaks in the energy range above 5.5~MeV for the six crystals is attributed to the crystal surface contributions from the decays of $^{226}$Ra~(sub-chain U3), $^{228}$Th~(sub-chain T2), and $^{227}$Th~(sub-chain A3), and maybe more dangerous than the internal contaminations that contribute to the grey area. Internal contributions from the decays of $^{227}$Th~(sub-chain A3) are dominant contributors in the energy range above 5.5~MeV, but their effect on the background in the ROI is negligible.
Overall, the weighted average of the fitted spectra is matched to the data. However, it does not match the spectral features of peaks in the data below 5~MeV. Moreover, the model does not explain some peaks in the data above 5.5~MeV, which is not shown in the model.
The detailed discussion on it will be continued in Sect.~\ref{sec:5.2}.

Based on the $\alpha$ background model for the six crystals, fitted activities of both the internal and surface are listed in Tables~\ref{tab:tabelfitbulk} and~\ref{tab:tabelfitsurface}. 
There are large contributions from the surface contaminations, while the fitted activities of most internal radionuclides are lower than the measured ones.
This is because the surface contaminants partially contribute to the full energy-deposition peaks as do the internal radionuclides.  
Peaks at 5979.3~keV and 6750.3~keV are attributed to the internal nuclides of $^{223}$Ra and $^{211}$Bi from the $^{227}$Th decay~(sub-chain A3), and they are comparable to the measured ones.

\subsection{Variable bins and systematics} 
\label{sec:5.2}
To understand the spectral features of peaks in the data, we compared alpha spectra obtained by generating surface contaminants at depths from 0 to 10~nm with a step size of 1~nm. We found that a few nm scale depth makes peaks at alpha energy due to energy loss by nuclear recoil escapes from the shallow depth and a deeper than those, like 10~nm, makes Q-value peaks. However, data shows peaks at around alpha energy or between alpha energy and Q-value depending on each crystal, which makes tens of keV difference from the model. Although we varied the depth at the nanoscale, it did not fit the peak shape of the data well and made peaks not well-matched at several energies. Knowing the crystal surface condition is necessary to explain whether it is caused by the surface condition or uncertainty induced by theoretical calculations used in the simulation model; we used the G4ScreenedNuclearRecoil class for the simulation of low-energy nuclear recoils from $\alpha$ surface contaminations, which provides good agreement with SRIM simulations, but not as precise as a few nanometer scales~\cite{screenednuclearrecoil}. Therefore, we simulated alpha energy spectra from the surface contaminants at zero depth to verify their contributions to the backgrounds without considering spectral distortion due to the surface contribution depending on the contamination depth. We varied bin sizes of the alpha spectra to cover spectral shape changes around Q-value peaks due to energy loss ranging from 0 to about 100~keV by nuclear recoil escapes from the surface and fitted the simulated distributions to the data. 
In the fit, we constrained activities of radioactive sources as fit parameters based on the measurements described in Sect.~\ref{sec:3} and the results of background modeling performed in Sect.~\ref{sec:5.1}.

\begin{figure}[t]
\centering
\includegraphics[width=0.45\textwidth]{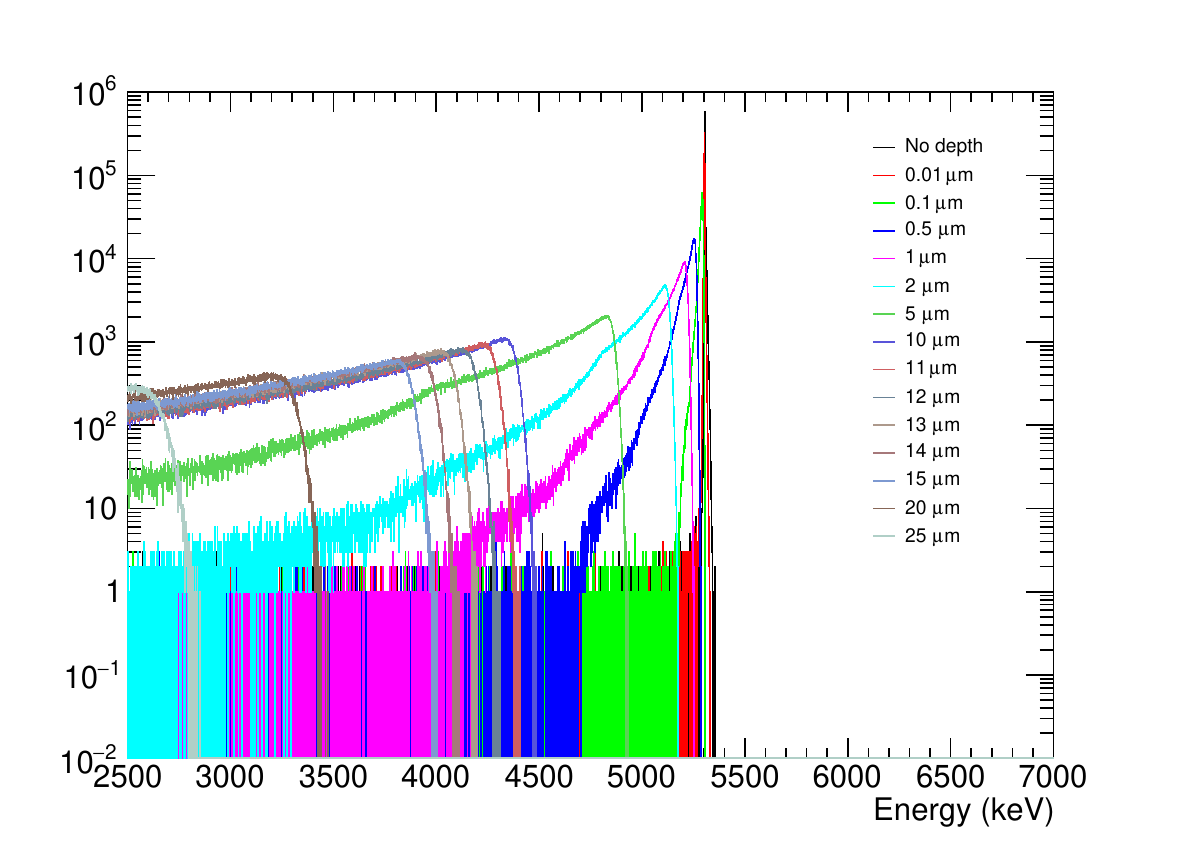}
\caption{
Simulated energy distributions of $^{210}$Pb events generated at the Vikuiti surface depths of 0, 0.1, 0.5, 1, 2, 5, 10, 15, 20, and 25~$\mu$m.
}
\label{fig:vikuitiPb210}
\end{figure}
\begin{figure*}[ht]
\begin{center}
\begin{tabular}{ccc}
\hspace*{0.25cm}\makebox[0.3\textwidth]{\includegraphics[width=0.3\paperwidth, height=0.23\textheight]{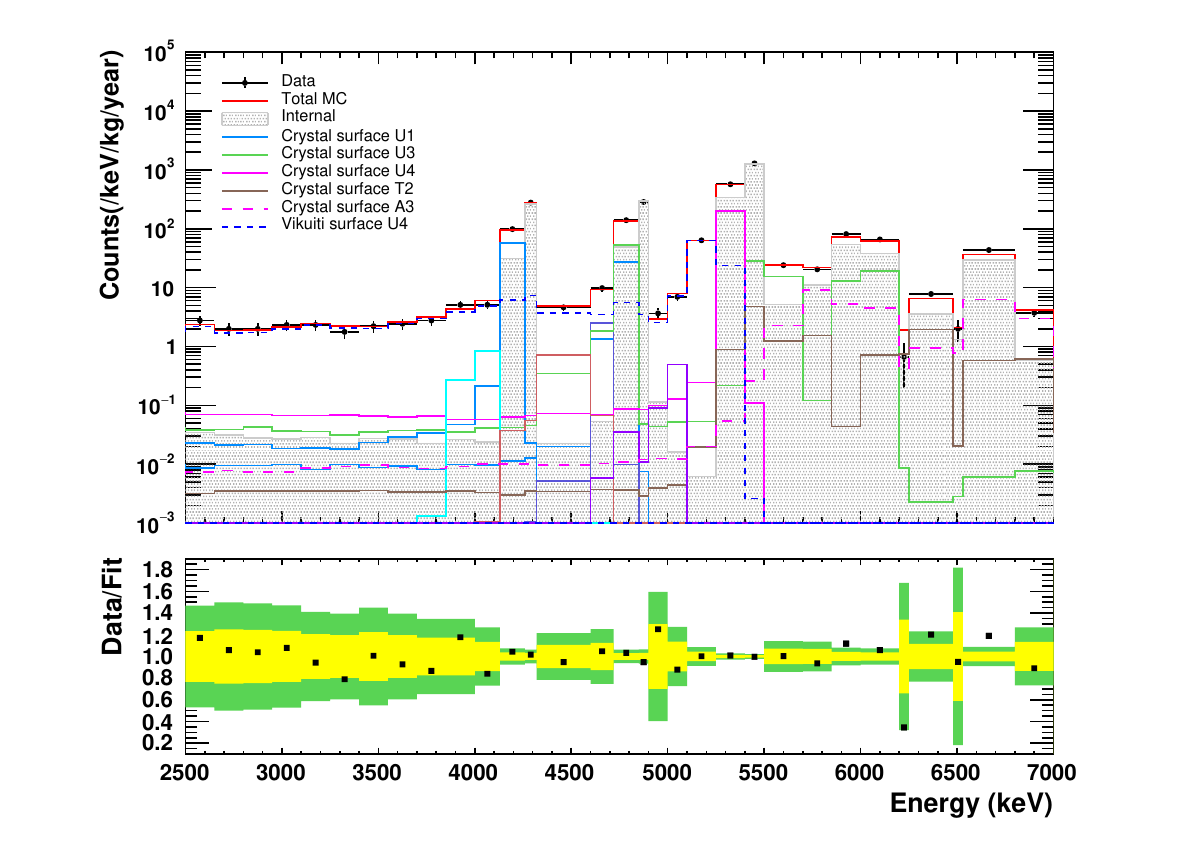}} &
\hspace*{0.25cm}\makebox[0.3\textwidth]{\includegraphics[width=0.3\paperwidth, height=0.23\textheight]{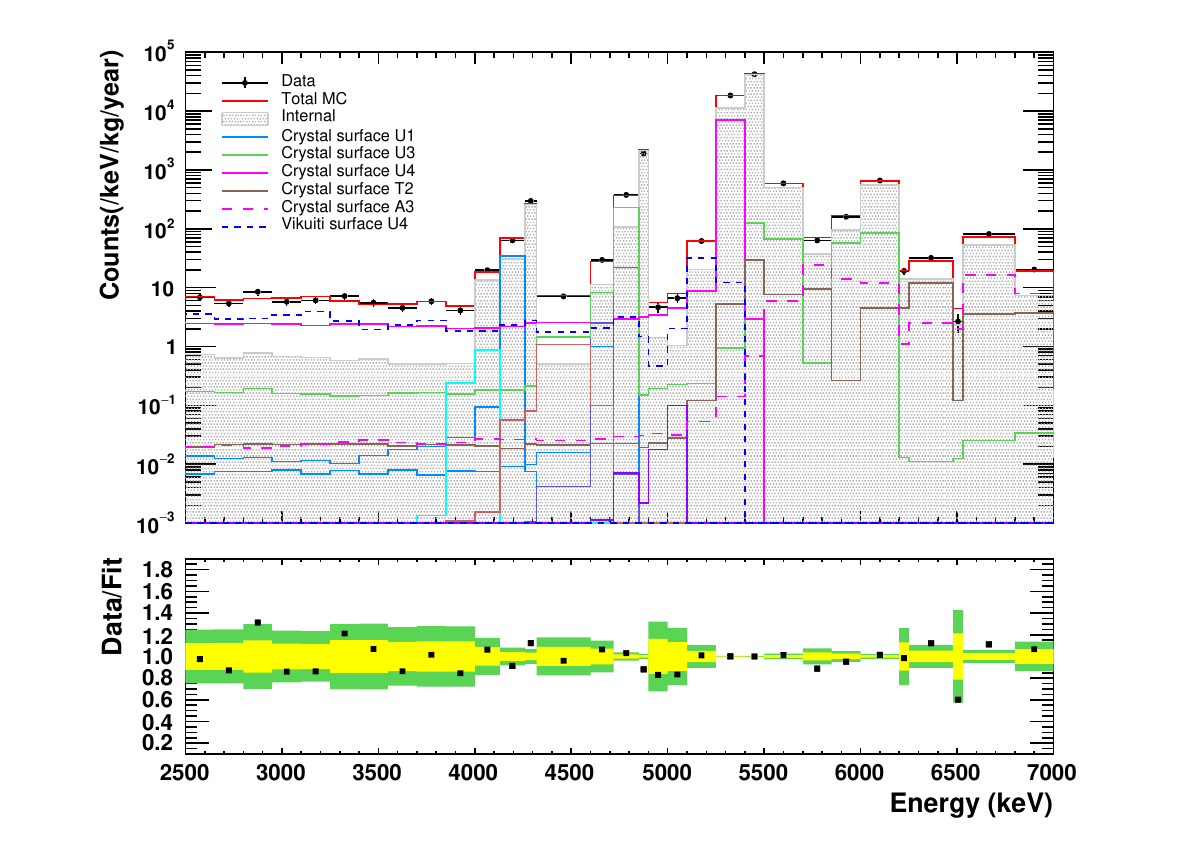}} &
\hspace*{0.25cm}\makebox[0.3\textwidth]{\includegraphics[width=0.3\paperwidth, height=0.23\textheight]{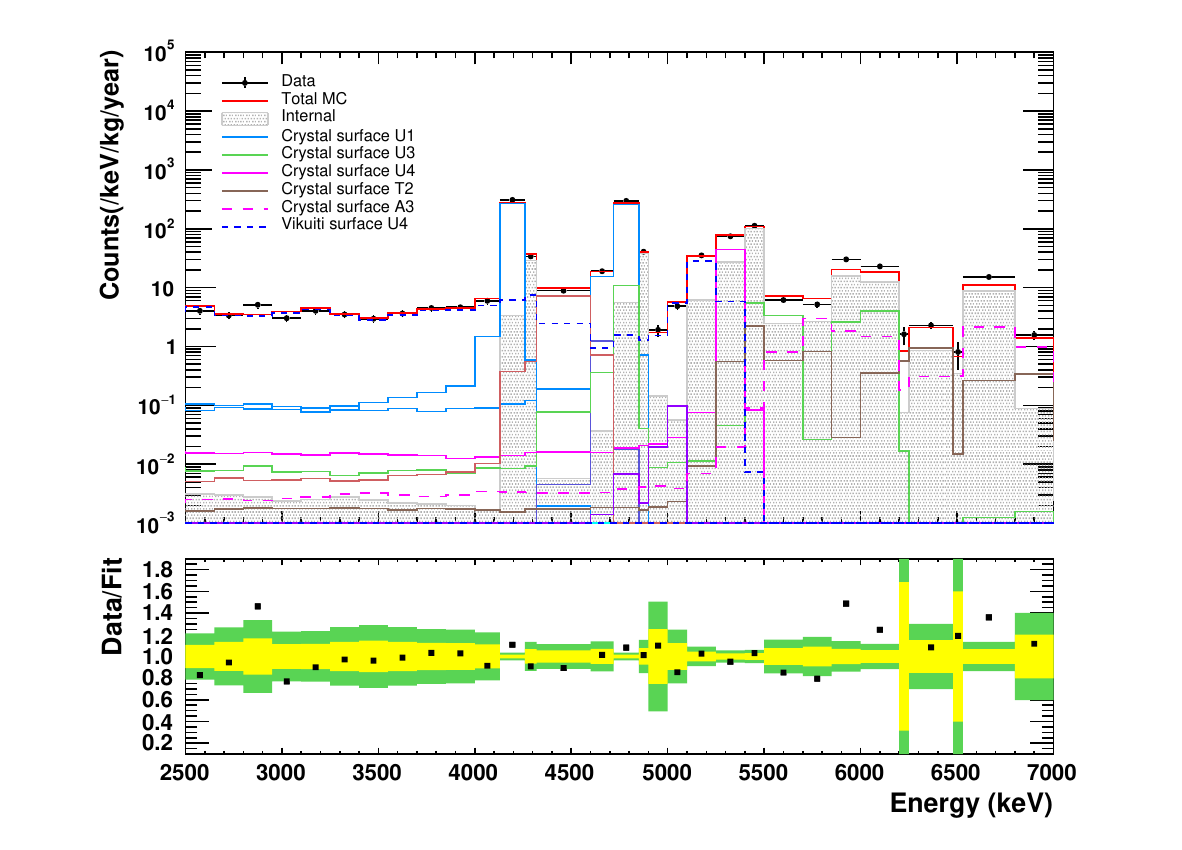}} \\
(a) C1 & (b) C2  & (c) C3 \\
\hspace*{0.25cm}\makebox[0.3\textwidth]{\includegraphics[width=0.3\paperwidth, height=0.23\textheight]{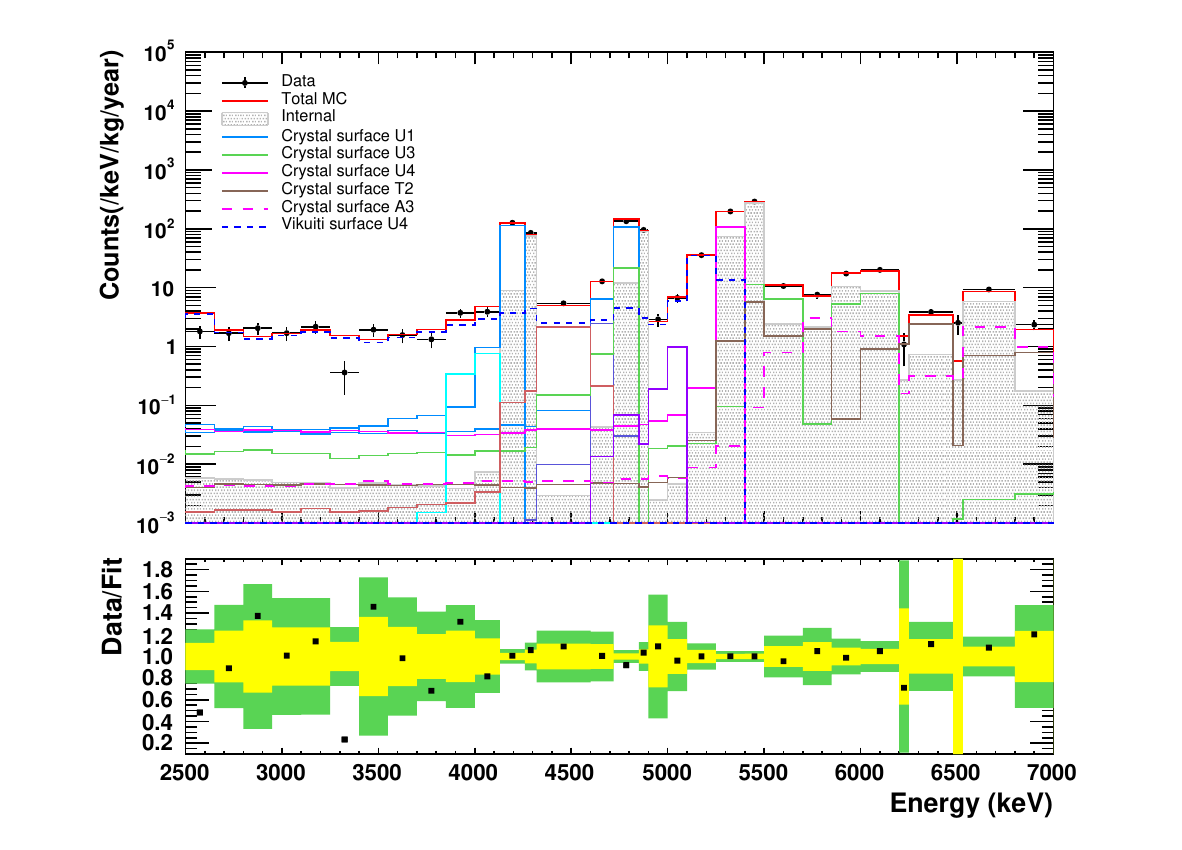}} &
\hspace*{0.25cm}\makebox[0.3\textwidth]{\includegraphics[width=0.3\paperwidth, height=0.23\textheight]{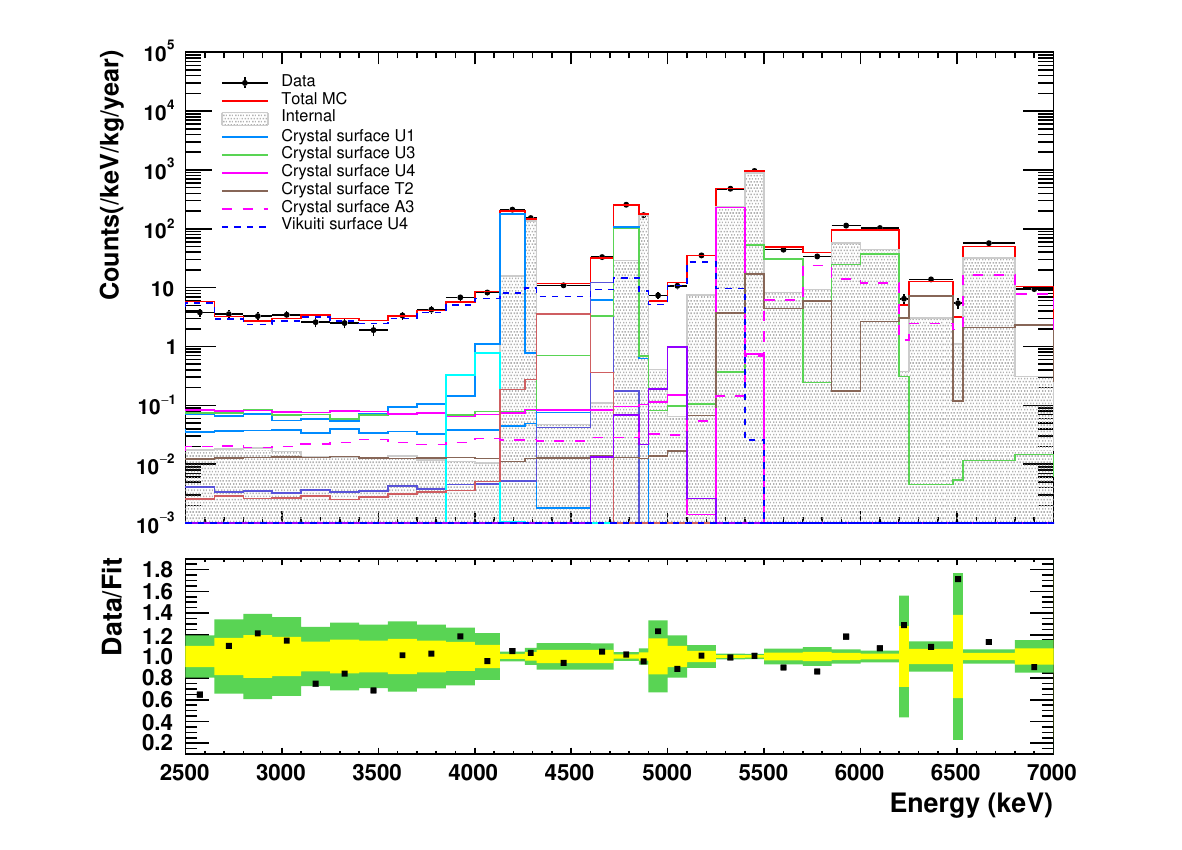}} &
\hspace*{0.25cm}\makebox[0.3\textwidth]{\includegraphics[width=0.3\paperwidth, height=0.23\textheight]{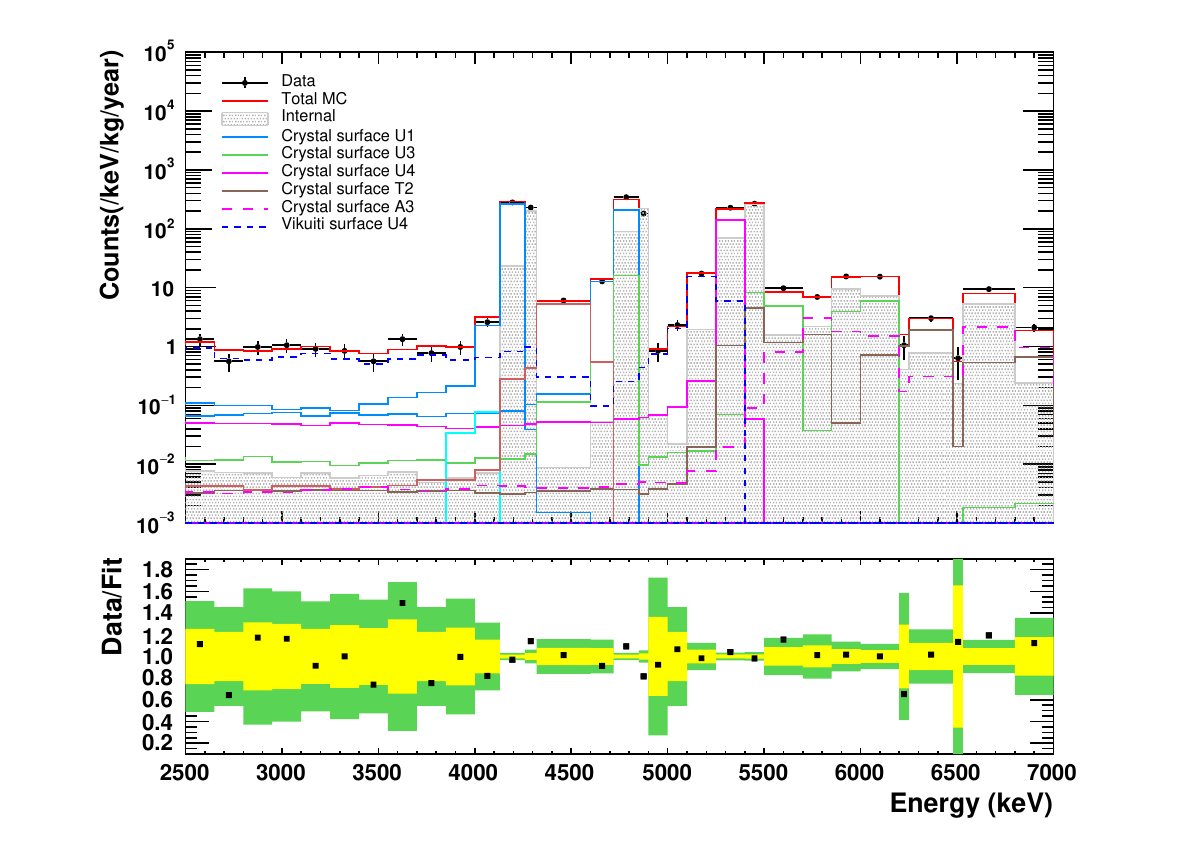}} \\
(d) C4 & (e) C5  & (f) C6 \\
\end{tabular}
\caption{ 
Measured energy spectra are compared with the fitted results for six crystals. The lower panel shows the bin by bin ratios of data to fitted results. The colored bands centered at 1 in the lower panel represent 1 and 2$\sigma$ of statistical uncertainty.
}
\label{fig:variablebinresults}
\end{center}
\end{figure*}

According to the $\alpha$ background model, the energy region below $\sim$5~MeV down to the continuum between 2.5 MeV and 4 MeV is mainly affected by the contribution of the surface $^{210}$Pb in the Vikuiti reflector film. At first, we considered two depths as shallow and deep contributions by generating contaminants randomly within those depths. Then, to improve their contributions in this area with variable bin sizes, we simulated alpha spectra at the depths such as 0, 0.1, 0.5, 1, 2, 5, 10, 15, 20, and 25~$\mu$m, as shown in Fig.~\ref{fig:vikuitiPb210}. They are included in the fit, and the other background sources used in the modeling described in Sect.~\ref{sec:5.1} are generated at zero depth to use in the fit, and as a result, we obtained spectra that matched better to the data.

Figure~\ref{fig:variablebinresults} shows the measured $\alpha$ energy spectra compared with the fitted results performed individually for each of the six crystals.
The blue dashed line is the sum of the fitted background spectra of $^{210}$Pb from various surface depths in the Vikuiti reflector film, which can provide the understanding of the depth profile of the Vikuiti surface  $^{210}$Pb contamination.

\subsection{Background in the ROI from surface contamination} 
\label{sec:5.3}
\begin{figure}[t]
\centering
\hspace*{0.2cm}\makebox[0.45\textwidth]{\includegraphics[width=0.45\paperwidth]{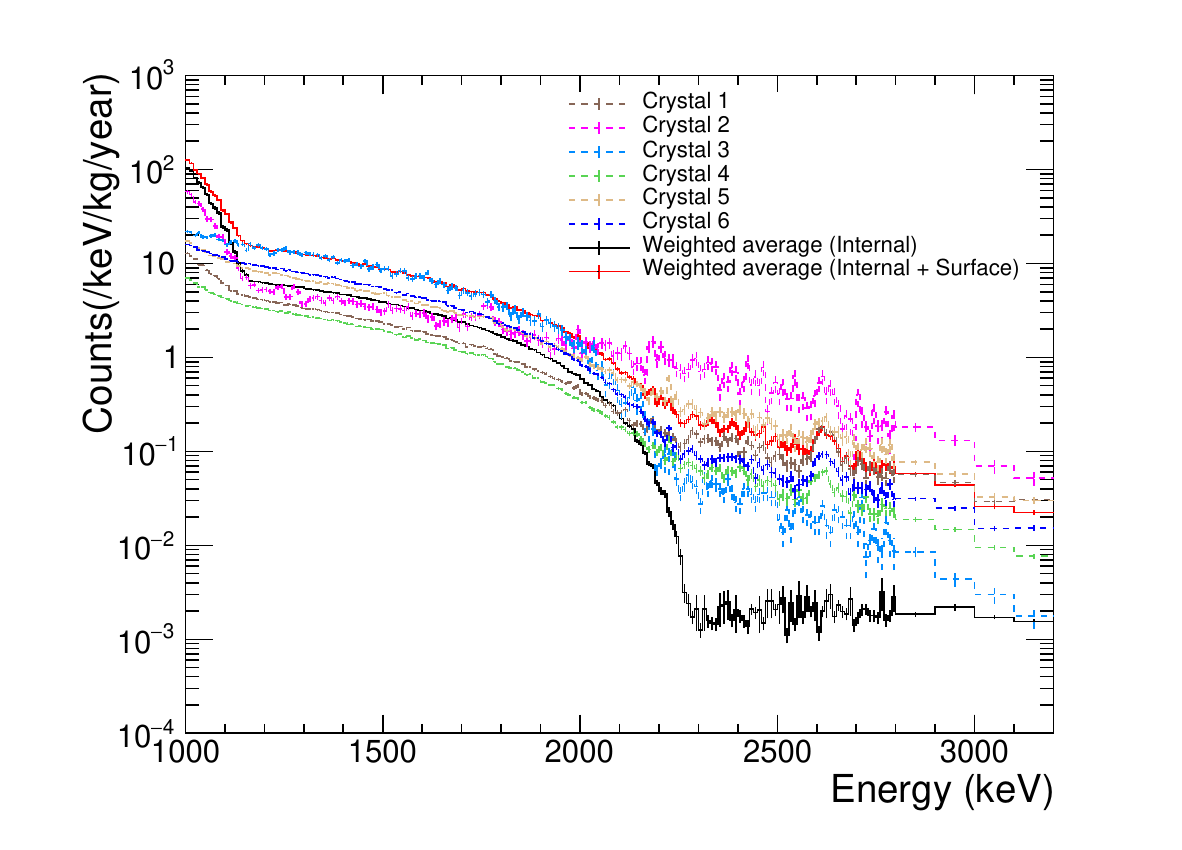}}
\caption{
Simulated energy distributions of $\beta/\gamma$ events from both internal and surface contaminations of the crystal detectors. Weighted averaged spectra from internal and surface~(red solid line), and internal only~(black solid line) contaminations are shown.
}
\label{fig:bginROI}
\end{figure}

Degraded $\alpha$ events from the crystal and Vikuti surfaces contribute to the continuum that extends down to the ROI region but they can be distinguished from 0$\nu\beta\beta$ events by the particle identification selection requirements described in Sect.~\ref{sec:3}. 
However, it is possible that $\beta$/$\gamma$ events are produced by the radioactive surface contamination as well.
Therefore, we estimated the background contributions in the 3.024$\sim$3.044~MeV ROI, which are produced by the surface contamination.
Figure~\ref{fig:bginROI} shows the simulated $\beta$/$\gamma$ energy spectra for each crystal that were obtained from the $\alpha$ modeling, which was described in the previous section. 
The background in the ROI due to the surface contamination is from the same isotopes as the internal contamination, i.e, $^{208}$Tl in T2 sub-chain and $^{214}$Bi in U3 sub-chain. $^{208}$Tl decay can be rejected by vetoing 30 minutes after the 6.2~MeV $\alpha$ precursor as in the case of internal contamination~\cite{luqman2017}. However, in the surface contamination case, the precursor $\alpha$ often does not give energy of 6.2~MeV, which results in lower veto efficiency. 
Other $\gamma$ emissions with $\sim$3~MeV energy come from the $^{214}$Bi decay in the $^{226}$Ra--$^{210}$Pb decay chain.
As shown in Fig.~\ref{fig:bginROI}, the ROI background level that is produced by the surface contamination is found to be  (3.02$\pm$0.46)$\times10^{-2}$~counts/keV/kg/yr~(ckky) for single-hit events averaged for all six detectors. The total background level in the ROI, from both internal and surface contaminations, is found to be (3.18$\pm$0.46)$\times10^{-2}$~ckky. 
The background levels averaged for the five detectors, excluding C2 are estimated to be (1.95$\pm$0.17)$\times10^{-2}$~ckky and (2.22$\pm$0.13)$\times10^{-4}$~ckky in the ROI for the surface and internal contributions, respectively.
The background level estimated for the surface contamination is higher than that of the AMoRE-II goal. 
Thus, a series of R\&D efforts with different surface conditions are being undertaken to avoid contributions from surface contamination.
%


\section{Conclusions}
We studied the surface background of CMO crystal detectors in AMoRE-Pilot using simulations with the Geant4 toolkit. 
To understand background contributions from both internal and surface radioactive contaminations in crystals and nearby materials, we simulated the decay chains of $^{238}$U, $^{232}$Th, and $^{235}$U for both internal and surface contaminations. The $\alpha$ background modeling was carried out by fitting the measured $\alpha$ spectra with the simulated data.
We found that the overall simulated background spectrum describes well the data for all six detectors in the 2.5~MeV--7.0~MeV energy region.  
Some features of the experimental spectrum can be reproduced by MC simulations for the surface contaminations of crystals and nearby materials. 
We found that Crystal 2, which was grown with single crystallization, has an exceptionally large internal contamination.  
The averaged background level, estimated at  (1.95$\pm$0.17)$\times10^{-2}$~ckky in the ROI, for the surface contaminations of the five crystals, is higher than that of the AMoRE-II goal.  Special efforts are needed to reduce these backgrounds in the future.
	
\section{Acknowledgement} 
This research was supported by: the Institute for Basic Science (Korea) under project code IBS-R016-D1 and IBS-R016-G1,  
the grant for the joint Korea-Ukraine project on AMoRE, and the National Research Foundation of Korea (NRF) grant funded by the Korean government (MSIT) (No. NRF-2018K1A3A1A13087769). 
This research was supported in part by the joint Ukraine-Republic of Korea R\&D project "Properties of neutrino and weak interactions in double beta decay of $^{100}$Mo", F.A. Danevich, V.V. Kobychev, N.V. Sokur and V.I. Tretyak acknowledge the National Research Foundation of Ukraine (Grant No. 2020.02/0011). 
Members of NRNU MEPhI group acknowledge the Ministry of Science and Higher Education of the Russian Federation, Project "Fundamental properties of elementary particles and cosmology" No 0723-2020-0041.
V.D. Grigoryeva, E.P. Makarov, and V.N. Shlegel acknowledge the Ministry of Science and Higher Education of the Russian Federation N121031700314-5.
M. B. Sari gratefully acknowledges the support from Directorate General of Higher Education of the Republic of Indonesia PMDSU scholarship.


\begin{thebibliography}{}

\bibitem {DellOro:2016tmg}
Stefano Dell’Oro, Simone Marcocci, Matteo Viel, Francesco Vissani, 
Adv. High Energy Phys. 2016, 2162659 (2016). \url{https://doi.org/10.1155/2016/2162659}
\bibitem{Mohapatra2007}
R. N. Mohapatra et al., 
Rept. Prog. Phys. 70 (2007) 1757–1867. \url{http://dx.doi.org/10.1088/0034-4885/70/11/R02} 
\bibitem{Patrignani2016}
P.A. Zyla et al. (Particle Data Group), Prog. Theor. Exp. Phys. 8 (2020) 083C01. \url{https://doi.org/10.1093/ptep/ptaa104}
\bibitem{Giunti2007}
C. Giunti, C. W. Kim, Fundamentals of Neutrino Physics and Astrophysics, Oxford, UK: Univ. Pr. (2007) 710 p, 2007. \url{https://doi.org/10.1093/acprof:oso/9780198508717.001.0001}

\bibitem{gando2016}
A. Gando, Y. Gando, T. Hachiya et al., Phys. Rev. Lett. 117, 082503 (2016). \url{https://doi.org/10.1103/PhysRevLett.117.082503}
\bibitem{agostini2020}
M. Agostini et al. (GERDA Collaboration), Phys. Rev. Lett. 125, 252502 (2020). \url{https://doi.org/10.1103/PhysRevLett.125.252502}
\bibitem{abgrall2014}
N. Abgrall, E. Aguayo, F.T. Avignone et al., Adv. High Energy Phys. 2014, 365432 (2014). \url{https://doi.org/10.1155/2014/365432}
\bibitem{alvis2019}
S. Alvis et al. (MAJORANA Collaboration), Phys. Rev. C 100, 025501 (2019). \url{https://doi.org/10.1103/PhysRevC.100.025501}
\bibitem{anton2019}
G. Anton et al. (EXO-200 Collaboration), Phys. Rev. Lett. 123, 161802 (2019). \url{https://doi.org/10.1103/PhysRevLett.123.161802}
\bibitem{adams2020}
D. Q. Adams et al. (CUORE Collaboration), Phys. Rev. Lett. 124, 122501 (2020). \url{https://doi.org/10.1103/PhysRevLett.124.122501}
\bibitem{azzolini2019}
O. Azzolini et al., Phys. Rev. Lett. 123, 032501 (2019). \url{https://doi.org/10.1103/PhysRevLett.123.032501}
\bibitem{arnold2021}
E. Armengaud, C. Augier, A. S. Barabash et al., Phys. Rev. Lett. 126, 181802 (2021). \url{https://doi.org/10.1103/PhysRevLett.126.181802}

\bibitem{Bhang2012}
H. Bhang et al., 
J. Phys. Conf. Ser. 375 (2012) 042023. \url{https://doi.org/10.1088/1742-6596/375/1/042023}
\bibitem{Alenkov2015}
V. Alenkov et al., 
\url{https://doi.org/10.48550/arXiv.1512.05957}
\bibitem{alenkov2019}
V. Alenkov et al., 
Eur. Phys. J. C (2019) 79:791. \url{https://doi.org/10.1140/epjc/s10052-019-7279-1}

\bibitem{luqman2017}
A. Luqman et al., 
Nucl. Instrum. Meth. Phys. Res., Sect. A 855 (2017) 140-147. \url{https://doi.org/10.1016/j.nima.2017.01.070}

\bibitem{bucci2009}
C. Bucci et al., 
Eur. Phys. J. A 41, 155-168 (2009). \url{https://doi.org/10.1140/epja/i2009-10805-7}
\bibitem{johnson2012}
R. A. Johnson et al., 
Nucl. Instrum. Meth. Phys. Res., Sect. A 693 (2012) 51-58. \url{https://doi.org/10.1016/j.nima.2012.06.043}
\bibitem{patavina}
L. Pattavina, PhD thesis, Univ. degli Studi di Milano-Bicocca and Univ. Claude Bernard Lyon 1 (2011).
\bibitem{alessandria}
F. Alessandria et al., 
Astropart. Phys. 45, 13-22 (2013). \url{https://doi.org/10.1016/j.astropartphys.2013.02.005}
\bibitem{yu2021}
G. H. Yu et al., Astropart. Phys. 126 (2021) 102518. \url{https://doi.org/10.1016/j.astropartphys.2020.102518}
\bibitem{vacri2021}
M. L. di Vacri, I. J. Arnquist, S. Scorza, E.W. Hoppe, J. Hall,
Nucl. Instrum. Meth. Phys. Res., Sect. A 994 (2021) 165051. \url{https://doi.org/10.1016/j.nima.2021.165051}

\bibitem{gbkim2017}
G. B. Kim et al., Astropart. Phys. 91 (2017) 105. \url{https://doi.org/10.1016/j.astropartphys.2017.02.009}

\bibitem{gbkim2016}
G. B. Kim, J. H. Choi, H. S. Jo, et al., IEEE Transactions on Nuclear Science 63(2), 539 (2016). \url{https://doi.org/10.1109/TNS.2015.2493529}
\bibitem{ikim2017}
I. Kim, H. S. Jo, C. S. Kang, et al., Supercond. Sci. Technol. 30(9), 094005 (2017). \url{https://doi.org/10.1088/1361-6668/aa7c73}

\bibitem{annenkov2008}
A. N. Annenkov et al, 
Nucl. Instrum. Methods Phys. Res. A584 (2008) 334. \url{https://doi.org/doi:10.1016/j.nima.2007.10.038}
\bibitem{kang2017}
C. S. Kang et al., 
Supercond. Sci. Technol. 30(8) (2017) 084011. \url{https://doi.org/10.1088/1361-6668/aa757a}

\bibitem{fomos}
JSC “Fomos Materials” \url{https://en.newpiezo.com/}
%

\bibitem{Tretyak2010}
V.I. Tretyak et al., Astropart. Phys. 33 (2010) 40. \url{https://doi.org/10.1016/j.astropartphys.2009.11.002}

\bibitem{gbkim2015}
G. B. Kim et al., Adv. High Energy Phys. 2015, 817530 (2015). \url{https://doi.org/10.1155/2015/817530}


\bibitem{Azzolini2021}
O. Azzolini et. al., Eur. Phys. J. C81 (2021) 722. \url{https://doi.org/10.1140/epjc/s10052-021-09476-z}

\bibitem{geant4}
S. Agostinelli et al., 
Nucl. Instrum. Meth. A506 (2003) 250-303. \url{https://doi.org/10.1016/S0168-9002(03)01368-8}
\bibitem{screenednuclearrecoil}
Marcus H. Mendenhall, Robert A. Weller,
Nucl. Instrum. Meth. B 227 (2005) 420–430. \url{https://doi.org/10.1016/j.nimb.2004.08.014}

\bibitem{HJLee2015}
H. J. Lee et al., 
Nucl. Instrum. Methods A 784 (2015) 508–512. \url{https://doi.org/10.1016/j.nima.2014.11.050}

\bibitem{SRIM}
James F. Ziegler, M. D. Ziegler, J. P. Biersack, 
Nucl. Instrum. Methods B 268 (2010) 1818-1823. \url{https://doi.org/10.1016/j.nimb.2010.02.091}

\bibitem{pdg} P.A. Zyla {\it et al.} (Particle Data Group), Prog. Theor. Exp. Phys. \textbf{2020} (2020) 083C01.
\url{https://doi.org/10.1093/ptep/ptaa104}

\bibitem{Azzolini19_background}
O. Azzolini et al.,
Eur. Phys. J. C 79 (2019) 583. \url{https://doi.org/10.1140/epjc/s10052-019-7078-8}

\bibitem{Sala2016}
E.Sala, I.S.Hahn, W.G.Kang, G.W.Kim, Y.D.Kim, M.H.Lee, D.S. Leonard, S. Y. Park, Journal of Physics Conference Series 718 (2016) 062050. \url{https://doi:10.1088/1742-6596/718/6/062050}
\bibitem{cosine2021_background}
G. Adhikari et al., Eur. Phys. J. C 81 (2021) 837. \url{https://doi.org/10.1140/epjc/s10052-021-09564-0}

\end{thebibliography}

\end{document}